\documentclass[10pt,journal,compsoc]{IEEEtran}

\ifCLASSOPTIONcompsoc
  \usepackage[nocompress]{cite}
\else
  \usepackage{cite}
\fi

\ifCLASSINFOpdf
\else
\fi
\usepackage{array}
\usepackage{marvosym}
\usepackage{tabularx}
\usepackage{amsmath}
\usepackage{graphicx}
\usepackage{algorithmic}
\usepackage{threeparttable}
\usepackage{titletoc}
\usepackage[ruled,linesnumbered]{algorithm2e}
\usepackage{cite}
\usepackage{amsfonts}
\usepackage{color}
\usepackage{colortbl}
\definecolor{mygray}{gray}{.9}
\usepackage{bm}
\usepackage{enumitem} 
\usepackage{hyperref}   

\definecolor{dark-green}{rgb}{0, 0.6, 0.25} 
\definecolor{citecolor}{rgb}{0,0.443,0.737} 
\definecolor{linkcolor}{rgb}{0.956,0.298,0.235} 
\hypersetup{
    colorlinks=true,
    linkcolor=linkcolor,
    filecolor=magenta,
    citecolor=dark-green,
    urlcolor=cyan 
}
\newcommand{\zhun}[1]{{\color{black}#1}}

\newcommand{\black}[1]{{\color{black}#1}}
\usepackage{booktabs}
\usepackage{caption}
\begin{document}
\title{
Modeling the Label Distributions for Weakly-Supervised Semantic Segmentation}
\author{Linshan Wu, Zhun Zhong, Jiayi Ma,~\IEEEmembership{Senior Member,~IEEE}, Yunchao Wei,~\IEEEmembership{Senior Member,~IEEE}, \par~Hao Chen,~\IEEEmembership{Senior Member,~IEEE}, Leyuan Fang\textsuperscript{\Letter},~\IEEEmembership{Senior Member,~IEEE}, and Shutao Li\textsuperscript{\Letter},~\IEEEmembership{Fellow,~IEEE}
        
\IEEEcompsocitemizethanks{\IEEEcompsocthanksitem Linshan Wu, Leyuan Fang, and Shutao Li are with the College of Electrical and Information Engineering, Hunan University, Changsha 410082, China.
E-mail: \href{mailto:linshanwu@hnu.edu.cn}{\black{linshanwu@hnu.edu.cn}}, \href{mailto:fangleyuan@gmail.com}{\black{fangleyuan@gmail.com}}, \href{mailto:shutao\_li@hnu.edu.cn}{\black{shutao\_li@hnu.edu.cn}}.
\IEEEcompsocthanksitem Zhun Zhong is with the Department of Information Engineering and Computer Science (DISI), University of Trento, Trento 38123, Italy. E-mail: \href{mailto:zhunzhong007@gmail.com}{\black{zhunzhong007@gmail.com}}.
\IEEEcompsocthanksitem Jiayi Ma is with the Electronic Information School, Wuhan University, Wuhan 430072, China. E-mail: \href{mailto:jyma2010@gmail.com}{\black{jyma2010@gmail.com}}.
\IEEEcompsocthanksitem Yunchao Wei is with the Institute of Information Science, Beijing Jiaotong University, Beijing 10044, China. E-mail: \href{mailto:yunchao.wei@bjtu.edu.cn}{\black{yunchao.wei@bjtu.edu.cn}}.
\IEEEcompsocthanksitem Hao Chen is with the Department of Computer Science and Engineering, The Hong Kong University of Science and Technology, Clear Water
Bay, Hong Kong. E-mail: \href{mailto:jhc@cse.ust.hk}{\black{jhc@cse.ust.hk}}.
}
\thanks{Corresponding author: Leyuan Fang and Shutao Li.}
}


\IEEEtitleabstractindextext{%
\begin{abstract}

Weakly-Supervised Semantic Segmentation (WSSS) aims to train segmentation models by weak labels, which is receiving significant attention due to its low annotation cost. Existing approaches focus on generating pseudo labels for supervision while largely ignoring to leverage the inherent semantic correlation among different pseudo labels. We observe that pseudo-labeled pixels that are close to each other in the feature space are more likely to share the same class, and those closer to the distribution centers tend to have higher confidence. Motivated by this, we propose to model the underlying label distributions and employ cross-label constraints to generate more accurate pseudo labels. In this paper, we develop a unified WSSS framework named Adaptive Gaussian Mixtures Model, which leverages a GMM to model the label distributions. Specifically, we calculate the feature distribution centers of pseudo-labeled pixels and build the GMM by measuring the distance between the centers and each pseudo-labeled pixel. Then, we introduce an Online Expectation-Maximization (OEM) algorithm and a novel maximization loss to optimize the GMM adaptively, aiming to learn more discriminative decision boundaries between different class-wise Gaussian mixtures. Based on the label distributions, we leverage the GMM to generate high-quality pseudo labels for more reliable supervision. Our framework is capable of solving different forms of weak labels: image-level labels, points, scribbles, blocks, and bounding-boxes. Extensive experiments on PASCAL, COCO, Cityscapes, and ADE20K datasets demonstrate that our framework can effectively provide more reliable supervision and outperform the state-of-the-art methods under all settings. Code will be available at \href{https://github.com/Luffy03/AGMM-SASS}{here}.

\end{abstract}

}

\maketitle
\IEEEdisplaynontitleabstractindextext
\IEEEpeerreviewmaketitle

\ifCLASSOPTIONcompsoc
\IEEEraisesectionheading{\section{Introduction}\label{sec:introduction}}
\else
\section{Introduction}
\label{sec:introduction}
\fi
\IEEEPARstart{S}{emantic} segmentation aims to assign the corresponding pixel-wise semantic labels for each given image, which is one of the most fundamental computer vision tasks. Modern deep learning based semantic segmentation models~\cite{FCN, Segnet, deeplab} trained with large amounts of manually labeled data have demonstrated outstanding achievements. However, the cost of collecting massive pixel-level annotations is particularly expensive, which heavily limits the development of semantic segmentation methods. Hence, numerous attempts~\cite{stc, A2GNN, online, LIID, salvage, EPS++} have been made toward Weakly-Supervised Semantic Segmentation (WSSS), which utilize different forms of weak labels for supervision, \emph{e.g.}, image-level label~\cite{MCIC, MCIS, wei_revisiting}, points~\cite{whatpoint, kernel_cut, seminar}, scribbles~\cite{PSI, Scribblesup, URSS}, and bounding box~\cite{boxsup, SDI, WSSL}.

\begin{figure}
	\centering
	\includegraphics[width=1\linewidth]{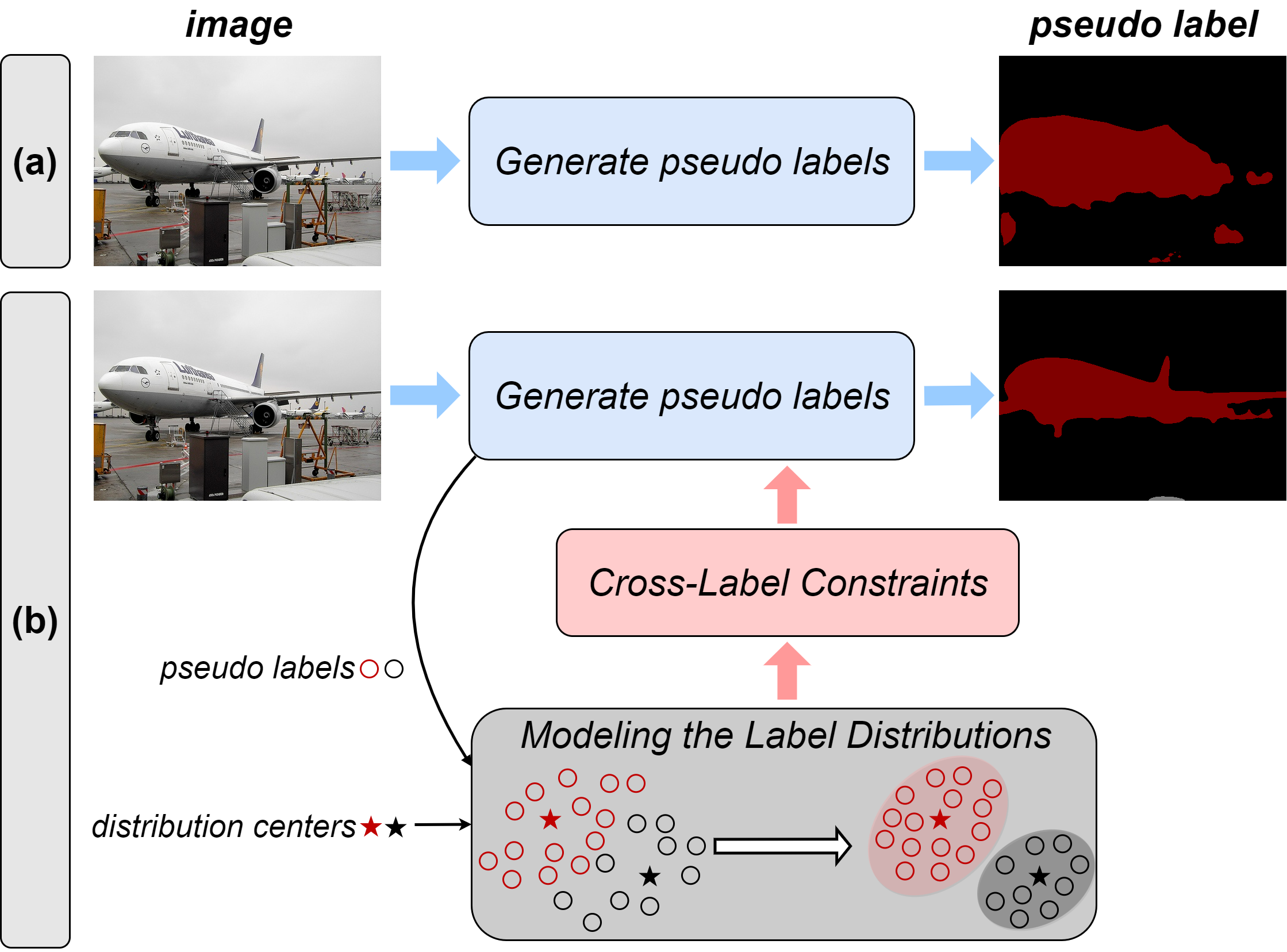}
	\caption{(a) Methods that generate pseudo labels by learning from initial weak labels (\emph{e.g.}, CAMs~\cite{cam, advcam}, points, scribbles, and boxes). (b) Our AGMM++ employs the cross-label constraints on the pseudo labels by modeling the label distributions.}
	\label{fig1}
\end{figure}

\begin{figure*}
	\centering
	\includegraphics[width=1\linewidth]{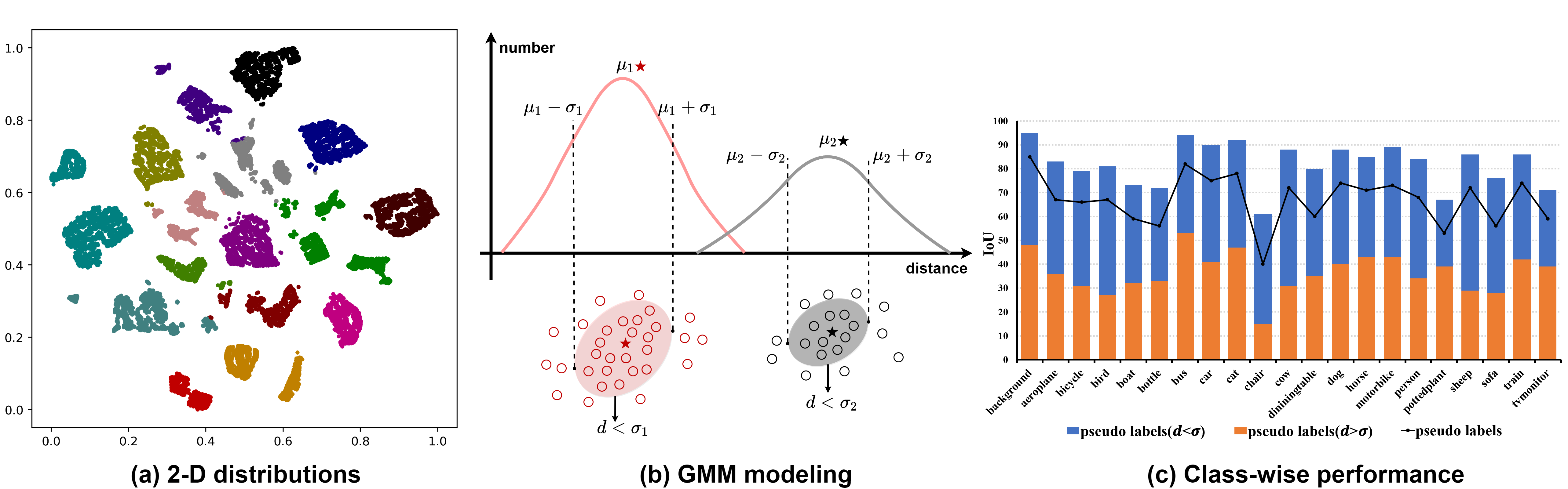}
	\caption{(a) Two-dimensional feature distributions visualization~\cite{t-sne} on the PASCAL dataset using a supervised baseline, which fits the Gaussian Mixture distributions. (b) We formulate a GMM to model the distributions of pseudo labels, where $\mu$ is the mean features, $\sigma$ is the variance, and $d$ represents the feature distance between each pseudo-labeled pixel and $\mu$. (c) We use the GMM to split the pseudo labels into two groups, \emph{i.e.}, $d<\sigma$ and $d>\sigma$. It can be seen that pseudo labels closer ($d<\sigma$) to the distribution centers have higher accuracy.}
	\label{fig2}
\end{figure*}



The main challenge in WSSS is to bridge the gap between weak labels and dense predictions. Existing methods~\cite{A2GNN, singlestage, random-walk, stc} mainly focus on generating pseudo labels from the initial weak labels for better supervision. Although some post-processing techniques\cite{A2GNN, PSA, random-walk, crf} can be used to refine the pseudo labels, these pseudo labels still inevitably encompass a lot of noise, which significantly hinders the training. To solve the problem, recent WSSS methods propose to introduce extra prior information for supervision~\cite{survey}, which can be roughly divided into three categories, \emph{i.e.}, cross-pixel similarity~\cite{PSA, AFA, SPML, IRN}, cross-image consistency~\cite{MCIC, MCIS, LIID, CISC_R, cian}, and cross-view consistency~\cite{l2g, toco}. Cross-pixel similarity indicates that pixels with similar low-level cues, \emph{e.g.}, color, brightness, and textures, probably belong to the same semantic class. Recent methods~\cite{TEL, kernel_cut, normcut, GridCut} further utilize the low-level affinity to regularize the final predictions. However, they generally ignore the gap between unreliable low-level visuals and high-level semantics. Cross-image and cross-view consistency further explore the feature consistency between different images and different views of the same image, respectively, aiming to learn consistent representations in the high-dimension feature space. However, they fail to directly supervise the final segmentation predictions. Thus, how to generate reliable supervision for unlabeled pixels remains to be the key problem in WSSS.

Previous methods~\cite{A2GNN,MCIC,MCIS,EPS++} only implicitly utilize the information of pseudo labels to supervise the segmentation model through the back-propagation of loss functions, while failing to explicitly leverage the high-level semantic correlation among different pseudo labels. In this paper, we propose that \zhun{leveraging the high-level cross-label relation can provide complementary information and assist in obtaining improved pseudo labels (as shown in Fig.~\ref{fig1}). This is inspired by the following observations. 
First, we find that the features of different pseudo labels fit the Gaussian Mixture distributions in the high-dimension space (Fig.~\ref{fig2}(a)). Based on this observation, we formulate a GMM to model the underlying distributions (Fig.~\ref{fig2}(b)). Additionally, we observe that pseudo-labeled pixels that are close to each other in the feature space are more likely to share the same class, and those closer to the distribution centers tend to have higher confidence (Fig.~\ref{fig2}(c)).} Motivated by this, we propose to model the label distributions by measuring the distance between each pseudo-labeled pixel and the distribution centers. Intuitively, if a pseudo label is closer to the distribution center of a class, it has higher confidence to be assigned as this class. In this way, we can easily associate the confidence of pseudo labels with the label distributions, providing extra information to generate more reliable pseudo labels.

In this paper, we propose a unified WSSS framework named Adaptive Gaussian Mixture Model, which is capable of solving different forms of weak labels. Our framework is equipped with a GMM to model the label distributions of each class in the high-dimension feature space, which can be easily incorporated into any traditional segmentation framework. Specifically, we first assign the initial weak labels as pseudo labels and calculate the feature distribution centers of pseudo-labeled pixels. Then, we build the GMM by adaptively measuring the distance between the centers and each pseudo-labeled pixel, where the confidences of pseudo labels are highly associated with their distance to distribution centers. In this way, we can leverage the GMM to refine the pseudo labels based on the label distributions, providing more reliable supervision. The process of GMM formulation works in an adaptive manner, where the parameters of GMM are dynamically adapted to the input features, achieving the end-to-end online self-supervision. In addition, the GMM is adaptively optimized during training, enabling us to learn more discriminative decision boundaries between different class-wise Gaussian mixtures adaptively.

This paper is an extension of our previous conference version AGMM~\cite{AGMM}. In this paper, we made significant and substantial modifications to AGMM, and we name the improved framework as AGMM++. \zhun{The new contributions of this paper}
include but are not limited to: \textbf{(1)} Compared to AGMM~\cite{AGMM} that can only tackle points and scribbles labels, the improved AGMM++ is a unified WSSS framework that is capable of solving different forms of weak labels, \emph{e.g.}, image-level labels, points, scribbles, blocks, and bounding-boxes. Specifically, we modify the formulation of GMM to handle different forms of weak labels. In addition, different from other unified WSSS frameworks~\cite{A2GNN, CDL, SPML}, our AGMM++ does not require a time-consuming multi-stage training process for pseudo labels generations. Instead, we propose an online approach for generating pseudo labels using GMM, enabling end-to-end training. \textbf{(2)} To effectively refine the generated pseudo labels, we further develop an Online Expectation–Maximization (OEM) algorithm to optimize the GMM during training, which significantly improves the accuracy of pseudo labels. \textbf{(3)} We conduct more extensive experiments with different backbones (ResNet~\cite{resnet} and ViT~\cite{vit}) on four different datasets, \emph{i.e.}, PASCAL~\cite{VOC}, COCO~\cite{COCO}, Cityscapes~\cite{city}, and ADE20K~\cite{ade20k}. 
\zhun{The experiments demonstrate that our AGMM++ consistently achieves state-of-the-art performance across various datasets and under different weakly-supervised settings. These results highlight the significant potential of our AGMM++ for widespread adoption in WSSS.} 
\textbf{(4)} We provide more detailed ablation studies and insightful analyses to demonstrate the effectiveness of the core components of our method. \zhun{These experiments further highlight the significance of modeling the label distributions, offering valuable insights that can inspire future research in the field of WSSS.} 


\begin{figure*}
	\centering
	\includegraphics[width=1\linewidth]{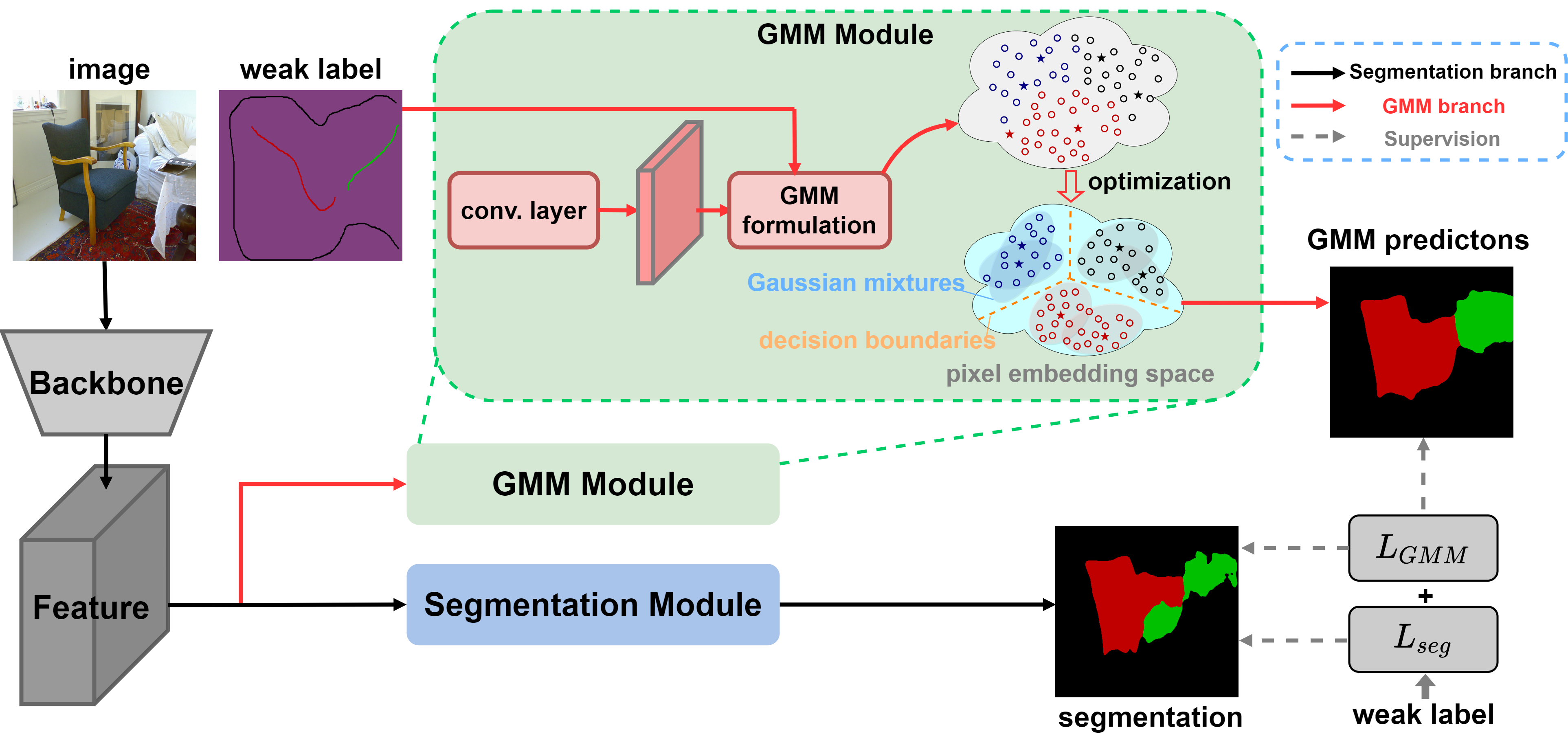}
	\caption{\textbf{The overall framework of AGMM++.} AGMM++ contains a segmentation module and a GMM module. The segmentation module is responsible for final segmentation predictions. Given an input image, the deep features extracted from the backbone are fed into the GMM module to generate soft GMM predictions as pseudo labels. Finally, we employ the segmentation and GMM predictions for online self-supervision. It is worth noting that the weak label can be points, scribbles, blocks, or coarse seed labels generated from image- or box-level labels. Here, we use scribbles as an example. The GMM module is discarded in the inference process.}
	\label{fig3}
\end{figure*}

\section{Related Work}
\subsection{Weakly-Supervised Semantic Segmentation}
To address the problem of expensive pixel-wise semantic segmentation annotations, \zhun{there has been a recent surge of study in the field of WSSS, resulting in numerous proposed methods.}
Various types of weak annotations have been widely explored, \emph{e.g.}, image-level label \cite{MCIC, MCIS, seam, wei_revisiting}, points~\cite{whatpoint, kernel_cut, seminar}, scribbles~\cite{PSI, Scribblesup, URSS}, and bounding box~\cite{boxsup, SDI, WSSL}. Among them, image-level label gains more attention due to less annotation cost. Existing WSSS methods based on image-level supervision rely on network visualization techniques, \emph{e.g.}, class activation maps (CAMs)~\cite{cam, gradcam}, which leverage localization maps from classifiers for supervision. However, they always struggle to make the localization maps capture more complete object contents. To alleviate this problem, MDC~\cite{wei_revisiting} proposes to apply dilated convolutions to extend the receptive fields. DSRG\cite{DSRG} proposes a seed region growing algorithm to expand more regions. FickleNet~\cite{ficklenet} utilizes random operators to force the model to activate more diverse parts. Although promising achievements are produced in these works, they are still heavily limited by the initial location maps and fail to provide reliable supervision for training a segmentation model.

Points, scribbles, and bounding boxes are stronger supervision signals since both class and localization information are provided. What's the Point~\cite{whatpoint} first uses point annotations to supervise a semantic segmentation model. ScribbleSup~\cite{Scribblesup} proposes to propagate scribble labels via a graphical model for supervision. Boxsup~\cite{boxsup} further exploits bounding boxes to supervise a semantic segmentation model. Most existing state-of-the-art WSSS methods~\cite{A2GNN, l2g, IRN, DCA, qiang} adopt a multi-stage pipeline for pseudo supervision, which generates pseudo labels from the initial weak labels, aiming to bridge the gap between weak supervision and dense predictions. Thus, the key challenge is to introduce extra reliable information to generate better pseudo labels. In the following, we briefly discuss three kinds of related approaches to address this problem.

\textbf{Cross-pixel similarity}: Cross-pixel similarity indicates that pixels with highly similar low-level cues, \emph{e.g.}, color, brightness, and texture, probably belong to the same semantic region in an image. This prior is derived from principles of perceptual grouping. Early WSSS methods~\cite{IRN, DSRG, random-walk, A2GNN, VoCo, BPG} propose to expand the initial weak labels (CAMs, points, scribbles, and bounding boxes) based on the similarity between different pixels. Specifically, IRN~\cite{IRN} utilizes the inter-pixel similarity on the attention maps to propagate the initial CAMs and generate more complete pseudo labels for supervision. RAWKS~\cite{random-walk} develops a random-walk label propagation algorithm based on pair-wise pixel similarity. In addition, many previous works~\cite{TEL, normcut, kernel_cut} also employ such low-level similarity to regularize the final segmentation predictions. However, they ignore the large gap between the low-level information and the high-level semantics, which provide extra unreliable supervision.

\textbf{Cross-image consistency}: Early WSSS methods generally only consider every single image and overlook the potential cross-image contexts. Some recent works~\cite{MCIC, MCIS, LIID, CISC_R, cian} propose that the cross-image consistency among a group of images can provide complementary information for each other to obtain better pseudo labels for supervision. Specifically, MCIS~\cite{MCIS} first develops a co-attention module to enforce the classifier to recognize the common semantics from co-attentive objects in a group of images, aiming to produce more complete pseudo labels for segmentation training. MCIC~\cite{MCIC} further proposes a memory bank mechanism and a transformer-based cross-image attention module to extract cross-image contexts from different images and store them as memory to generate better pseudo labels. The main idea is to extract shared features from different images and learn discriminative CAMs as pseudo labels for further supervision. However, these generated pseudo labels are still limited by the ability of the initial classifiers, inevitably containing a lot of noise.

\textbf{Cross-view consistency}: Cross-view consistency, referring to the same object showing
consistency in different views, is another commonly used prior in WSSS~\cite{l2g, toco}. L2G~\cite{l2g} proposes a local-to-global knowledge transfer framework to contrast common objects from global views and local views. ToCo~\cite{toco} further proposes a transformer-based consistent learning method to contrast global images and local patch tokens extracted by transformer. Although consistency learning based methods enable the model to learn more discriminative features, they still fail to directly supervise the final predictions in the category-level~\cite{TEL}. 

These previous methods mainly rely on exploring the feature similarity from the original images to refine the pseudo labels, while largely ignoring the already-existed semantic correlations among different pseudo labels. In this paper, we propose to explicitly leverage the cross-label relation to employ constraints on the pseudo label generations. Specifically, we propose to model the label distributions and measure the distance between different pseudo labels and the distribution centers. In this way, we can associate the accuracy of pseudo labels with the label distributions, thus producing better pseudo labels based on the cross-label relation. 

\subsection{Gaussian Mixture Model}

Gaussian Mixture Model (GMM) is a typical probabilistic model for representing mixture distributions. GMM has been widely applied to model the distributions of hand-crafted features in an unsupervised way~\cite{gmm1, gmm2, gmm3, gmm4, gmm5, grabcut}. Previous methods propose to use Expectation–Maximization (EM) algorithms~\cite{em1, em2, em3} to formulate GMMs, which demand initial human-interactive prior estimates and iterative parameter updates. Instead, in WSSS, the initial weak labels (CAMs, points, scribbles, and boxes) are available. Since we always learn a segmentation model by borrowing knowledge from the weak labels, it is conceivable to regard these initial weak labels as accurate prior information for GMM formulation. Thus, we can easily formulate a GMM with the help of the annotated information. In this paper, we leverage information of weak labels and employ an effective self-supervision loss function to adaptively optimize the GMM. In addition, the EM algorithm is also adopted to refine the pseudo labels in an online manner. 

\section{Method}

\subsection{Overall Framework}\label{sec3_1}

The overall framework of our proposed AGMM++ is illustrated in Fig.~\ref{fig3}, which contains a GMM module to generate pseudo labels for supervision and a segmentation module to produce segmentation predictions. Similar to previous settings~\cite{TSCD, toco, A2GNN, CDL}, the initial weak labels are used to supervise the segmentation predictions. In our AGMM++, we further leverage the GMM module to generate pseudo labels based on the initial weak labels. 

Specifically, given an input image, the segmentation module directly predicts segmentation results $P$ for joint supervision with the initial weak labels $y$ and GMM predictions $G$. In the GMM module, we first feed the extracted deep features from the backbone into a convolution layer to squeeze the number of channels, aiming to save the computation costs of the GMM formulation. Then, we formulate a GMM and generate GMM predictions $G$ as pseudo labels, providing extra supervision for the segmentation predictions $P$. Specifically, since the formulation of GMM is derivable, we do not stop the gradient of the GMM module during the training process. Thus, the GMM is also adaptively optimized during training, aiming to learn more discriminative decision boundaries between different class-wise Gaussian mixtures. \zhun{This optimization process allows us to generate more accurate pseudo labels.} 
It is worth noting that since the GMM module is discarded in the inference process, our AGMM++ does not introduce extra computation costs during testing.

\begin{figure}
	\centering	\includegraphics[width=1\linewidth]{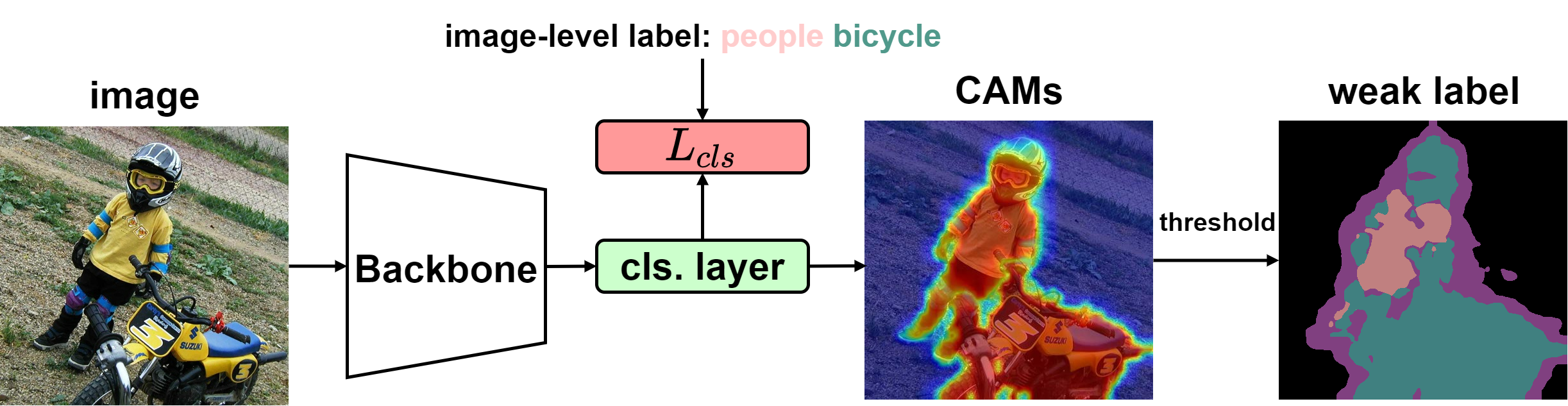}
	\caption{An example of generating weak labels from image-level label. A classification layer is used to output classification scores as CAMs~\cite{cam}, which is supervised with $L_{cls}$ using the class labels. Then, we use a threshold to transform CAMs to initial weak labels. In this paper, we follow the ToCo~\cite{toco} settings to generate CAMs.}
	\label{cam}
\end{figure}

Our AGMM++ is a unified WSSS framework that can handle different types of initial weak labels, including image-level labels, points, scribbles, blocks, and bounding-boxes. For image-level labels, we follow the common manners~\cite{toco, AFA, singlestage} to generate pixel-wise labels by online CAM generation, as illustrated in Fig.~\ref{cam}. It is worth noting that the CAMs are generated jointly with the training process of AGMM++, enabling us to achieve an end-to-end training procedure. For sparse annotations, \emph{i.e.}, points, scribbles, and blocks, they are completely accurate to provide completely reliable supervision. However, for weak labels generated from image- and box-level labels, they inevitably contain a lot of noise. Following the previous works~\cite{EPS++, MCIC, MCIS, toco}, we also assign these noisy weak labels as the initial pseudo labels for supervision. Then, we use the GMM module to refine these pseudo labels online, achieving more reliable supervision.



\subsection{GMM Formulation}\label{sec3_2}

Given an input image with $K$ annotated classes, we build a GMM with $K$ Gaussian mixture components. In GMM formulation, the first step is to calculate the centers ${\mu}_{i}$ of distribution. Previous methods~\cite{em2, em3, grabcut} generally initialize the centers randomly and update them with iterative optimization. In WSSS, since we learn the segmentation model by borrowing the knowledge from initial weak labels, it is conceivable to assign the weak labels as the initial distribution centers.

Different from the previous version of AGMM~\cite{AGMM}, we further use a convolution layer to squeeze the channels of deep features extracted from the backbone. The purpose is to reduce the computation cost of GMM formulation. Specifically, for the $i_{th}$ Gaussian mixture component, we first calculate the mean features of pseudo-labeled pixels $x_{li}$ belonging to $i_{th}$ class as the mean ${\mu}_{i}$:
\begin{equation}\label{mu}
	{\mu}_{i} = \frac{1}{|x_{li}|}\sum_{{\forall}{x}{\in}{x_{li}}}f(x, \theta),
\end{equation}
where $f(x, \theta)$ are the extracted deep features, $\theta$ are the learnable parameters of model. As stated in Section~\ref{sec3_1}, we do not stop the gradients of GMM module during training. Thus, $\theta$ is also adaptively updated through the back-propagation of the proposed loss functions during training.

Once obtaining the mean features ${\mu}_{i}$, the variance ${\sigma}_{i}$ of the $i_{th}$ component can be calculated as:
\begin{equation}\label{sigma}
{\sigma}_{i} = \sqrt{\frac{1}{|x_{li}|}\sum_{\forall{x}{\in}{x_{li}}}d^{2}},
\end{equation}
where $d$ measures the feature distance between each pixel in the $i_{th}$ component and the distribution centers ${\mu}_{i}$: 
\begin{equation}\label{distance}
d = \frac{1}{|C|}\sum_{{\forall}{c}{\in}{C}}f(x, \theta)-{\mu}_{i},
\end{equation}
where we calculate the mean distance along the channel dimension for deep features $f(x, \theta)$ with $C$ channels. Then, we build a GMM to model the label distributions. With the GMM, we produce the GMM predictions $G$ as pseudo labels for supervision:
\begin{equation}\label{gmm}
	G = \sum_{i}^{K}g_{i}(x, \theta, {\mu}_{i}, {\sigma}_{i}) = \sum_{i}^{K}{e}^{-\frac{d^2}{2{\sigma}_{i}^2}}.
\end{equation}
It is worth noting that compared to the typical GMM, we discard the regularization term $\frac{1}{\sqrt{2{\pi}{\sigma}^{2}}}$. In this way, we can guarantee that for each class, the GMM prediction scores $g_{i}$ range from 0 to 1, enabling us to conduct self-supervision with the segmentation predictions $P$. These GMM predictions $G$ are in the form of soft scores, denoting the probabilistic of each pixel belonging to which class-wise Gaussian mixture.

\textbf{Remark}. Our AGMM++ works in an adatpive manner from two perspectives. First, the parameters of GMM are dynamically adapted to the input features, achieving end-to-end online self-supervision. Specifically, the parameters of the GMM, \emph{i.e.}, number of components $K$, mean $\mu$, and variance $\sigma$, are dynamically adapted to the input images. Thus, our AGMM++ can dynamically generate reliable GMM predictions for different input images, enabling us to conduct online supervision. Second, the GMM module is adaptively optimized during training, enabling us to learn more discriminative decision boundaries between different class-wise Gaussian mixtures. \zhun{In the experiments, we will demonstrate the effectiveness of our adaptive method compared to non-adaptive baselines.}

\textbf{Discussion}. In our previous version of AGMM~\cite{AGMM}, we only consider sparse labels (points and scribbles) for WSSS training. Since these sparse annotations are completely accurate, it is appropriate to assign these sparse labels to calculate the distribution centers. However, aiming to solve different forms of weak labels in our improved AGMM++, we also assign the coarse weak labels (\emph{e.g.}, CAMs) as pseudo labels to formulate GMM. These weak labels only cover partial regions and inevitably contain a lot of noise. In this case, it is not appropriate to assign these noisy labels to calculate the distribution centers. To alleviate the impact of these coarse and partial weak labels, we further introduce an Online Expectation Maximization (OEM) algorithm to refine the GMM predictions online, as described in Section~\ref{sec3_3}.

\begin{figure}
	\centering
    \includegraphics[width=1\linewidth]{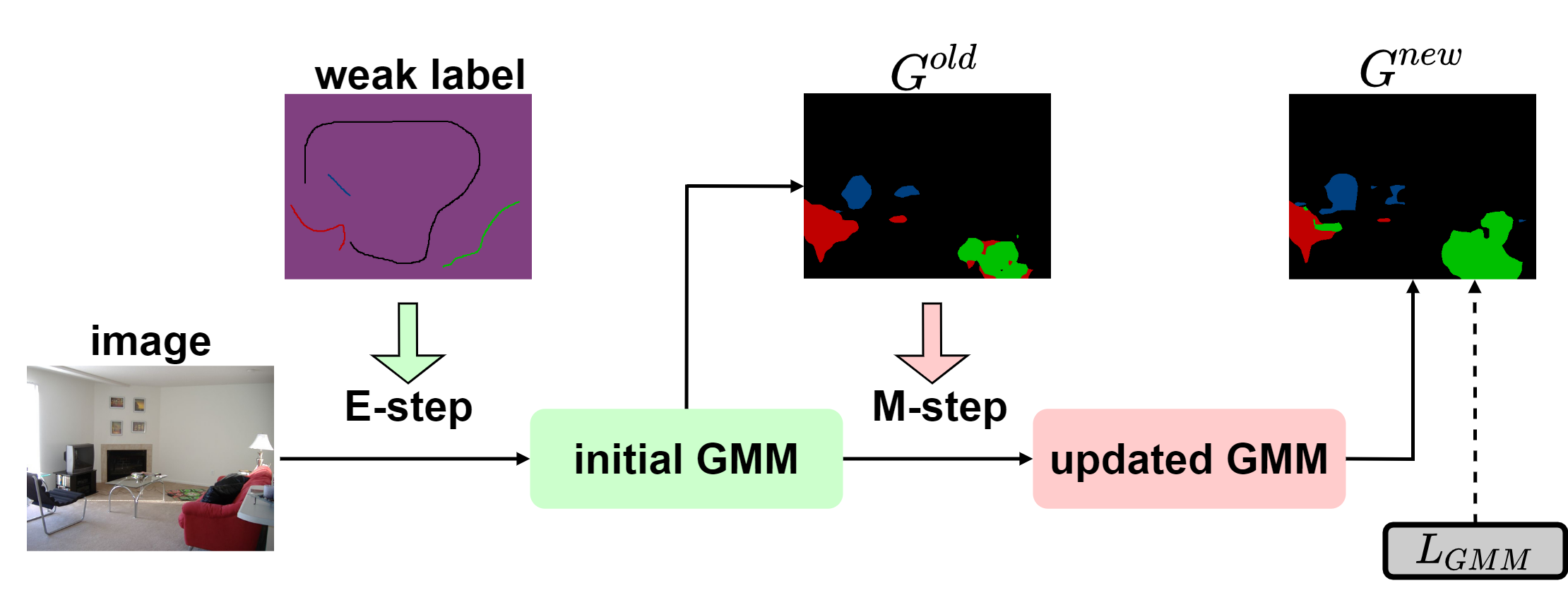}
    	\caption{The process of OEM algorithm. E-step and M-step represent the Expectation and Maximization steps, respectively. $L_{GMM}$ is used to supervise $G^{new}$.}
	\label{em}
\end{figure}

\subsection{Online Expectation Maximization}\label{sec3_3}

\zhun{The Online Expectation Maximization (OEM) algorithm is designed to refine the GMM predictions online. Similar to typical EM algorithm~\cite{em1, em2, em3}, OEM also contains two steps, an expectation step (E-step) and a maximization step (M-step) (as shown in Fig.~\ref{em}). However, in comparison to the typical EM algorithm, we have made several modifications to our OEM to tailor it specifically for the WSSS framework.}

\textbf{E-step:} According to Eqs.~\ref{mu} and \ref{sigma}, we first build an initial GMM by assigning the weakly-labeled pixels as the distribution centers. Here, the initial means $\mu$, variance $\sigma$, and pseudo labels $G$ are re-named as $\mu^{old}$, $\sigma^{old}$, and $G^{old}$, respectively. Then, the $G^{old}$ is used in the M-step to obtain updated GMM $G^{new}$.

\textbf{Remark}. Compared to the traditional EM algorithm~\cite{em1, em2, em3}, OEM does not require to randomly initialize the parameters and estimate posterior probabilities, since we can leverage the information of weak labels to formulate the initial GMM.

\textbf{M-step:} After obtaining $G^{old}$, the $G^{old}$ is further input to update the GMM as follows:
\begin{equation}\label{mu_new}
	{\mu}^{new}_{i} = \frac{1}{|G^{old}_{i}|}\sum_{{\forall}{x}{\in}{G^{old}_{i}}}f(x, \theta),
\end{equation}
\begin{equation}\label{sigma_new}
	{\sigma}^{new}_{i} = \sqrt{\frac{1}{|G^{old}_{i}|}\sum_{\forall{x}{\in}{G^{old}_{i}}}d^{2}},
\end{equation}
where $d$ is formulated similar to Eq.~\ref{distance}:
\begin{equation}\label{distance_new}
d = f(x, \theta)-{\mu}^{new}_{i},
\end{equation}
which measures the distance between each pseudo-labeled pixel produced by $G^{old}$ and the new centers ${\mu}^{new}$. Thus, the re-generated pseudo labels $G^{new}$ can be formulated as:
\begin{equation}\label{gmm_new}
	G^{new} = \sum_{i}^{K}g^{new}_{i}(x, \theta, {\mu}^{new}_{i}, {\sigma}^{new}_{i}) = \sum_{i}^{K}{e}^{-\frac{d^2}{2{{\sigma}^{new}_{i}}^2}}.
\end{equation}

\textbf{Remark}. Compared to the initial GMM $G^{old}$ that assigns weak labels to calculate the parameters, the updated GMM $G^{old}$ further assigns the generated pseudo labels $G^{old}$ for formulations. The motivation is that the weak labels only cover partial areas in the input images, as shown in Fig.~\ref{em}. Specifically, for points and scribbles, $x_{li}$ only covers less than 0.1\% pixels of the input images, which makes it difficult to estimate accurate GMM parameters. Compared to these sparse labels, $G^{old}$ covers all of the pixels in the image, which can provide more complete information. In addition, compared to the weak labels generated from image-level labels, $G^{old}$ are more complete and accurate. Thus, with the M-step, we can further re-estimate more accurate parameters for $G^{new}$. 

The M-step aims to optimize the GMM, where the goal of optimization can be formulated as follows:
\begin{equation}\label{opti_goal}
	\theta^{*},\mu^{*},\sigma^{*} = \arg\min-\sum_{i}^{K}\log[g(x, \theta, {\mu}, {\sigma})],
\end{equation}
which is equivalent to maximizing the conditional likelihood. The purpose of optimization is to re-estimate the GMM and make it fit better to the distributions. More specifically, Eq.~\ref{opti_goal} aims to encourage each pixel into one specific class-wise Gaussian mixture and learn more discriminative decision boundaries. \zhun{It is important to highlight that rather than employing  time-consuming iterative EM methods~\cite{em1, em2, em3} to refine the Gaussian Mixture Model (GMM), we design a maximization loss function to optimize the GMM online, which is both effective and efficient. We next will introduce the details of loss functions of our AGMM++ in Section~\ref{sec3_4}.}



\subsection{Training with AGMM++}\label{sec3_4}

\textbf{Segmentation Loss}. In WSSS, the input pixels $x$ can be separated into two parts: weakly-labeled pixels $x_{l}$ and unlabeled pixels $x_{u}$. As for the labeled pixels $x_{l}$, following the previous works~\cite{EPS++, MCIC, MCIS, toco}, we assign the corresponding weak labels $y_{l}$ as pseudo labels for supervision with a partial cross-entropy loss. The segmentation loss $L_{seg}$ is defined as follows:
\begin{equation}\label{L_seg}
	L_{seg} = -\frac{1}{|y_l|}\sum_{{\forall}{y}{\in}{y_l}}[{y}log(P_{i})+(1-y)log(1-P_{i})],
\end{equation}
where $P$ is the segmentation predictions. As for the unlabeled pixels $x_{u}$, there is no available label for supervision. In this paper, we generate GMM predictions as pseudo labels to provide supervision for the unlabeled pixels $x_{u}$.

\textbf{GMM Loss}. The GMM loss $L_{GMM}$ contains three parts, \emph{i.e.}, a self-supervision loss $L_{self}$, a weak loss $L_{weak}$, and a contrastive loss $L_{con}$. First, given the GMM predictions $G$, we assign them for self-supervision with the segmentation predictions $P$. We adopt a cross-entropy form to formulate the self-supervision loss function $L_{self}$ as follows:
\begin{equation}\label{L_self}
	L_{self} = -\frac{1}{|x|}\sum[Glog(P) + (1-G)log(1-P)].
\end{equation}
In the initial process of training, the accuracy of $G$ is much higher than $P$. With collaborative optimization of segmentation and GMM modules, the accuracy of $P$ and $G$ are both progressively improved. The details are described in the experiments of Sec.~\ref{sec4_2_1}.

Then, we also assign the weak labels $y_{l}$ to supervise $G$, the weak loss $L_{weak}$ is defined as follows:
\begin{equation}\label{L_weak}
	L_{weak} = -\frac{1}{|y_l|}\sum_{{\forall}{y}{\in}{y_l}}[{y}log(G) + (1-y)log(1-G)],
\end{equation}
where the $L_{weak}$ is used to bridge the gap between the initial weak labels $y_{l}$ and the generated pseudo labels $G$. 

In addition, to learn discriminative decision boundaries between different Gaussian mixtures, we further design a contrastive loss $L_{con}$. In our previous version AGMM~\cite{AGMM}, we only consider the distance between different centers and define $L_{con}$ as:
\begin{equation}\label{L_con_old}
	L_{con} = \frac{2}{K(K+1)}\sum_{{\forall}{i, j}{\in}{K}, i{\neq}j}{e}^{-({\mu}_{i} - {\mu}_{j})^2}.
\end{equation}
In our improved AGMM++, we aim to encourage each pixel into a specific Gaussian mixture, which can employ stronger supervision to each pixel. Thus, we re-define the $L_{con}$ as:
\begin{equation}\label{L_con}
	L_{con} = \frac{1}{N^{+}}\sum_{i=j}1-{e}^{-({f}_{i} - {\mu}_{j})^2}+\frac{1}{N^{-}}\sum_{i{\neq}j}{e}^{-({f}_{i} - {\mu}_{j})^2},
\end{equation}
where $N^{+}/N^{-}$ counts the number of positive/negative pairs, ${f}_{i}$ represents the deep features of a pixel that belongs to the $i_{th}$ component. 

Equipped with these three loss functions, we employ the GMM predictions $G$ and the segmentation predictions $P$ to supervise each other mutually. With weights $\lambda$, the total loss function $L_{GMM}$ for GMM predictions $G$ can be summarized as follows:
\begin{equation}\label{L_gmm}
	L_{GMM} = \lambda_{s}L_{self} + \lambda_{w}L_{weak} + \lambda_{c}L_{con},
\end{equation}
where $\lambda_{s}$, $\lambda_{w}$, and $\lambda_{c}$ are the weights of loss functions.

\textbf{Overall Loss}. In summary, the overall loss function $L$ in our AGMM++ framework is formulated as follows:
\begin{equation}\label{L_all}
	L = \lambda_{seg}L_{seg} + \lambda_{G}L_{GMM},
\end{equation}
where $\lambda_{seg}$ and $\lambda_{G}$ are the weights of $L_{seg}$ and $L_{GMM}$, respectively.

In addition, as stated in Fig.~\ref{cam}, for WSSS based on image-level labels, the loss function for classification $L_{cls}$ should be also considered. Thus, the overall loss function $L$ will be:
\begin{equation}\label{L_all_cls}
	L = \lambda_{c}L_{cls} +\lambda_{seg}L_{seg} + \lambda_{G}L_{GMM}.
\end{equation}

\begin{figure}
	\centering
	\includegraphics[width=1\linewidth]{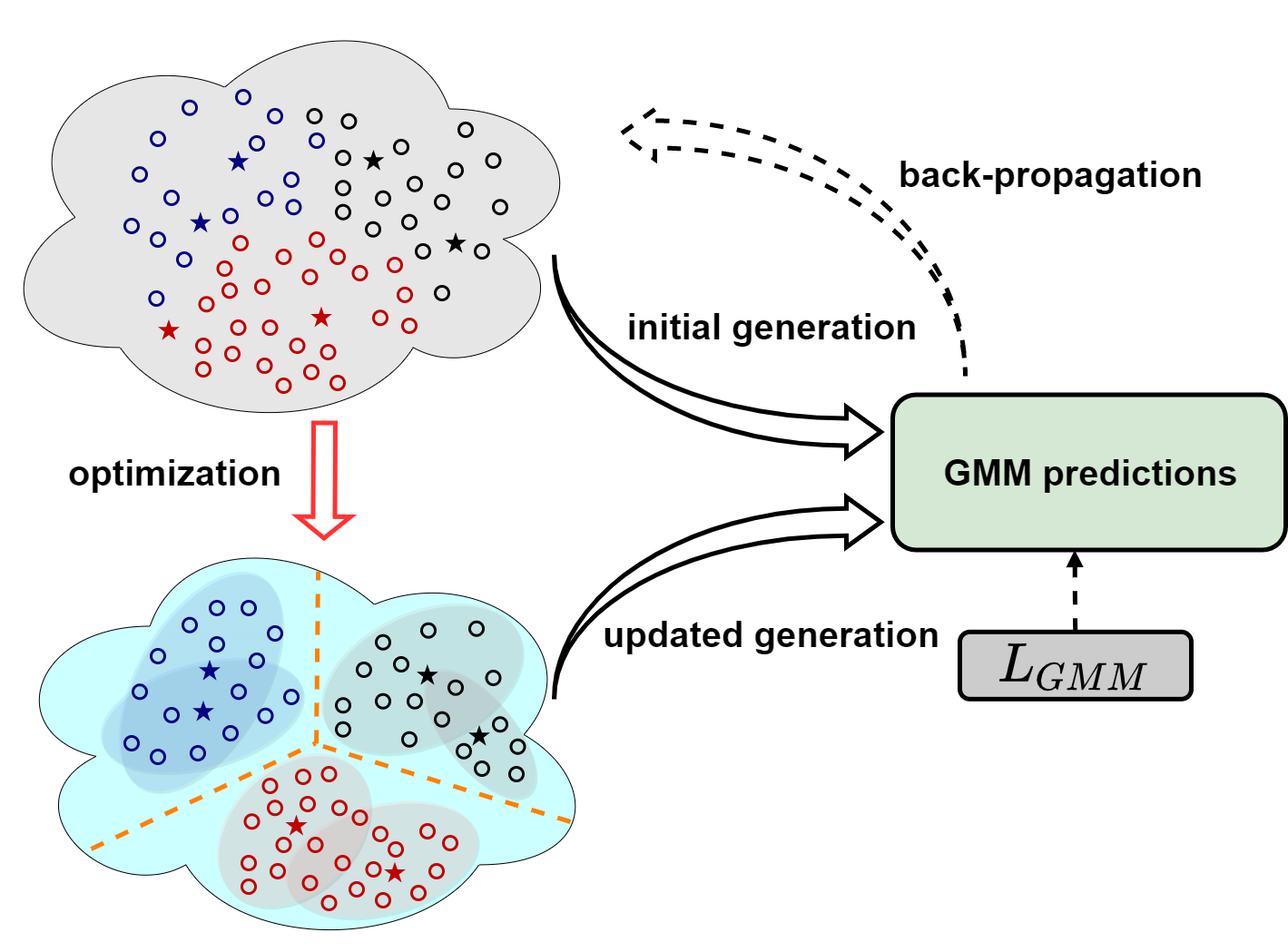}
	\caption{The optimization process of our GMM predictions. With the proposed loss functions according to Eq.~\ref{L_gmm}, we adaptively learn more discriminative category-wise Gaussian mixtures during the optimization process.}
	\label{gmm_opti}
\end{figure}

\textbf{Remark.} Our probabilistic GMM predictions $G$ are also optimized progressively during the back-propagation of $L_{GMM}$, as shown in Fig.~\ref{gmm_opti}. In our AGMM++ framework, with our proposed $L_{GMM}$, we assign each pixel to a specific Gaussian mixture, guiding us to employ strong supervision to each pixel. In addition, AGMM++ pulls the different Gaussian mixtures of different classes from each other, enabling us to learn more discriminative category-wise representations. In this way, we can use the GMM to generate more accurate pseudo labels for supervision. 

\section{Experiments}

To verify the effectiveness of our proposed method, we conduct extensive experiments on four widely-used semantic segmentation datasets: PASCAL VOC 2012~\cite{VOC}, COCO~\cite{COCO}, Cityscapes~\cite{city}, and ADE20K~\cite{COCO} datasets for different weakly-supervised settings, including image-level, point-level, scribble-level, box-level, and block-level supervised settings. The datasets and implementation details are described in the supplementary material. In this section, we first report the results of our proposed method compared with state-of-the-art methods. We then perform extensive ablation experiments for our approach. Finally, qualitative visualization results are presented to further demonstrate the effectiveness of our method.


\subsection{Comparison with State-of-the-Art Methods}
In this section, we compare our AGMM++ with existing state-of-the-art methods under different WSSS settings on different datasets. 

\begin{table}
	\setlength{\abovecaptionskip}{0.pt}
	\setlength{\belowcaptionskip}{-0.em}
	\centering
	\footnotesize
 \caption{Experimental results of the WSSS methods with image-level labels on the Pascal dataset. MT represents multi-stage training.}
\begin{threeparttable}
	\begin{tabular}{l|c|c|c|c}
		\toprule[1.2pt]
		\textbf{Method}
		&\textbf{Publications}
		&\textbf{Backbone} &\textbf{Val}
		&\textbf{Test}\\
		\hline
        \multicolumn{5}{c}{\textbf{Single-stage WSSS methods}}\\
        \hline
		RRM~\cite{RRM} &AAAI20 &R38 &62.6 &62.9\\
        1stage~\cite{singlestage} &CVPR22 &R38 &62.7 &64.3\\
		AFA~\cite{AFA} &CVPR22 &MiT-B1 &66.0 &66.3\\
        SLRNet~\cite{SLRNet} &IJCV22 &R38 &67.2 &67.6\\
        TSCD~\cite{TSCD} &AAAI23 &MiT-B1 &67.3 &67.5\\
        ToCo~\cite{toco} &CVPR23 &ViT-B &69.8 &70.5\\
        \hline
        \rowcolor{mygray}
        AGMM++ &- &R101 &68.0 &69.4\\
        \rowcolor{mygray}
        AGMM++ &- &ViT-B &\textbf{71.5} &\textbf{71.7}\\
        \hline
        \multicolumn{5}{c}{\textbf{Multi-stage WSSS methods}}\\
        \hline
		SEAM~\cite{seam} &CVPR20 &R38 &64.5 &65.7\\
		AdvCAM~\cite{advcam} &CVPR21 &R101 &68.1 &68.0\\
		OC-CSE~\cite{OCCSE} &ICCV21 &R38 &68.4 &68.2\\
		CPN~\cite{CPN} &ICCV21 &R38 &67.8 &68.5\\
		RIB~\cite{RIB} &NIPS21 &R101 &68.3 &68.6\\
		VWL~\cite{VWL} &IJCV22 &R101 &69.2 &69.2\\
        SIPE~\cite{SIPE} &CVPR22 &R101 &68.8 &69.7\\
		AMN~\cite{AMN} &CVPR22 &R101 &69.5 &69.6\\
		ReCAM~\cite{RECAM} &CVPR22 &R101 &68.4 &68.2\\
		ADELE~\cite{ADELE} &CVPR22 &R101 &69.3 &68.8\\
        ESOL~\cite{ESOL} &NIPS22 &R101 &69.9 &69.3\\
        CDL~\cite{CDL} &IJCV23 &R101 &71.1 &72.4\\
        BECO~\cite{beco} &CVPR23 &R101 &71.8 &71.8\\
        \hline
        \rowcolor{mygray}
        AGMM++(MT) &- &R101 &72.3 &72.4\\
        \rowcolor{mygray}
        AGMM++(MT) &- &ViT-B &\textbf{73.0} &\textbf{73.1}\\
		\toprule[1.2pt]
	\end{tabular}
    \end{threeparttable}
\label{pascal_img}
\end{table}%

\begin{table}
	\setlength{\abovecaptionskip}{0.pt}
	\setlength{\belowcaptionskip}{-0.em}
	\centering
	\footnotesize
 \caption{Experimental results of the WSSS methods with image-level labels on the COCO dataset.}
\begin{threeparttable}
	\begin{tabular}{l|c|c|c}
		\toprule[1.2pt]
		\textbf{Method}
		&\textbf{Publications}
		&\textbf{Backbone} &\textbf{Val}\\
		\hline
        \multicolumn{4}{c}{\textbf{Single-stage WSSS methods}}\\
        \hline
		AFA~\cite{AFA} &CVPR22 &MiT-B1 &38.9\\
        SLRNet~\cite{SLRNet} &IJCV22 &R38 &35.0\\
        TSCD~\cite{TSCD} &AAAI23 &MiT-B1 &40.1\\
        ToCo~\cite{toco} &CVPR23 &ViT-B &41.3\\
        \hline
        \rowcolor{mygray}
        AGMM++ &- &R101 &40.7\\
        \rowcolor{mygray}
        AGMM++ &- &ViT-B &\textbf{42.5}\\
		\hline
        \multicolumn{4}{c}{\textbf{Multi-stage WSSS methods}}\\
        \hline
		SEAM~\cite{seam} &CVPR20 &R38 &31.9\\
		CONTA~\cite{conta} &NIPS20 &R101 &32.8\\
		EPS~\cite{EPS} &CVPR21 &R101 &35.7\\
		CDA~\cite{CDA} &ICCV21 &R38 &33.2\\
		OC-CSE~\cite{OCCSE} &ICCV21 &R38 &36.4\\
		RIB~\cite{RIB} &NIPS21 &R101 &40.5\\
        VWL~\cite{VWL} &IJCV22 &R101 &36.2\\
		WOoD~\cite{WOOD} &CVPR22 &R38 &36.2\\
		MCTformer~\cite{MCTFormer} &CVPR22 &R38 &42.0\\
        SIPE~\cite{SIPE} &CVPR22 &R38 &43.8\\
		URN~\cite{URN} &AAAI22 &R101 &40.5\\
        ESOL~\cite{ESOL} &NIPS22 &R101 &42.6\\
        \hline
        \rowcolor{mygray}
        AGMM++(MT) &- &R101 &43.8\\
        \rowcolor{mygray}
        AGMM++(MT) &- &ViT-B &\textbf{44.6}\\
        \toprule[1.2pt]
	\end{tabular}
    \end{threeparttable}
\label{coco_img}
\end{table}%

\subsubsection{Image-level Supervised WSSS}
We first conduct WSSS experiments with image-level labels on PASCAL VOC and COCO datasets, as shown in Tables~\ref{pascal_img} and \ref{coco_img}, respectively. All the results are obtained after multi-scale testing and CRF post-processing~\cite{deeplab}. We divide the existing methods into two groups, \emph{i.e.}, single-stage and multi-stage training. The multi-stage methods~\cite{seam, advcam, RIB, beco} require a multi-stage training (MT) process to generate pseudo labels. They can achieve better results but require extra training burden. Single-stage methods~\cite{RRM, singlestage, AFA, toco, SLRNet} generate pseudo labels online, which are more efficient. It is worth noting that we can also adopt the multi-stage training (MT) strategy to further improve the accuracy of single-stage methods. To conduct fair comparisons with single-stage methods~\cite{RRM, singlestage, AFA, toco, SLRNet} and multi-stage methods~\cite{seam, advcam, RIB, beco}, we conduct AGMM++ with and without MT in the experiments. Specifically, \zhun{when adapting MT, we use the pseudo labels generated in the first stage to train a new model in the second stage.}

As shown in Table~\ref{pascal_img}, when using ViT-B as the backbone, our proposed AGMM++ achieves 71.5\% and 71.7\% mIoU on the PASCAL val and test sets, respectively, which largely outperform all previous single-stage methods by a clear margin. Specifically, \zhun{under the single-stage setting, our AGMM++ gains 1.7\% mIoU improvements over the existing best method ToCo~\cite{toco}.}
%
In addition, without using MT, our AGMM++ also obtains competitive results compared with the multi-stage methods. Moreover, with using MT, AGMM++ achieves extra improvements, outperforming all existing multi-stage methods.

\zhun{As shown in Table~\ref{coco_img}, when using ViT-B as backbone and without using MT, our AGMM++ achieves the state-of-the-art performance with 42.5\% mIoU on COCO val set. In addition, by using MT, AGMM++ achieves 44.6\% mIoU, which clearly demonstrates the effectiveness of our method.}

\begin{table}
	\setlength{\abovecaptionskip}{0.pt}
	\setlength{\belowcaptionskip}{-0.em}
	\centering
	\footnotesize
 \caption{Experimental results of the point-supervised WSSS methods on the Pascal validation set. Experimental settings with backbone, multi-stage training, and CRF post-processing are also considered.}
\begin{threeparttable}
	\begin{tabular}{l|c|c|c|c|c}
		\toprule[1.2pt]
		\textbf{Method}
		&\textbf{Pub.}
		&\textbf{Back.} &\textbf{Mul.}
		&\textbf{CRF} &\textbf{mIoU}\\
		\hline
		What Point~\cite{whatpoint} &ECCV16 &VGG16 &- &- &43.4\\
        KernelCut~\cite{kernel_cut} &ECCV18 &R101 &$\checkmark$ &$\checkmark$ &57.0\\
        SEAM~\cite{seam} &CVPR20 &R38 &$\checkmark$ &$\checkmark$ &66.3\\
        A2GNN~\cite{A2GNN} &PAMI21 &R101 &$\checkmark$ &$\checkmark$ &66.8\\
        Seminar~\cite{seminar} &ICCV21 &R101 &$\checkmark$ &$\checkmark$ &72.5\\
        SPML~\cite{SPML} &ICLR21 &R101 &- &$\checkmark$ &73.2\\
		DBFNet~\cite{DBFNet} &TIP22 &R101 &- &- &66.8\\
		TEL~\cite{TEL} &CVPR22 &R101 &- &- &64.9\\
        TSCD+SEAM~\cite{TSCD} &AAAI23 &MiT-B1 &$\checkmark$ &$\checkmark$ &70.2\\
		AGMM~\cite{AGMM} &CVPR23 &R101 &- &- &69.6\\
        \hline
        \rowcolor{mygray}
        AGMM++ &- &R101 &- &- &72.3\\
        \rowcolor{mygray}
        AGMM++ &- &R101 &- &$\checkmark$ &74.1\\
        \rowcolor{mygray}
        AGMM++ &- &ViT-B &- &- &75.6\\
        \rowcolor{mygray}
        AGMM++ &- &ViT-B &- &$\checkmark$ &\textbf{77.8}\\
		\toprule[1.2pt]
	\end{tabular}
    \end{threeparttable}
\label{pascal_point}
\end{table}%

\subsubsection{Point-Supervised WSSS}
We conduct point-supervised WSSS experiments on PASCAL and Cityscapes datasets. The results on PASCAL are reported in Table~\ref{pascal_point}.
In the previous point-supervised WSSS methods, TEL~\cite{TEL} produces the best performance with 64.9\% mIoU under consistent settings. Specifically, TEL~\cite{TEL} uses the tree filter methods~\cite{tree1, tree2, tree3} to model low-level affinity for regularization. Compared with TEL~\cite{TEL}, our previous work AGMM~\cite{AGMM} achieves 69.6\% mIoU with the same settings. Our improved version AGMM++ achieves 73.3\% mIoU and outperforms the previous AGMM~\cite{AGMM} by 2.7\%. We also conduct experiments with ViT-B~\cite{vit} and CRF post-processing. It can be seen that our AGMM++ can achieve 77.8\%, which outperforms existing state-of-the-art methods by a large margin.

The results on the Cityscapes dataset are shown in Table~\ref{table:city_point}. Following the settings of AGMM~\cite{AGMM}, we employ ResNet-50~\cite{resnet} and DeeplabV3+~\cite{deeplabv3} as the backbones. For fair comparisons, we also conduct the experiments with several existing point-supervised WSSS methods~\cite{TEL, seminar, normcut}. Since the Cityscapes dataset contains more complex scenes with diverse objects and cluttered backgrounds, the low-level affinity is not obvious in the Cityscapes dataset. Thus, the low-level regularization methods~\cite{TEL, normcut} cannot achieve obvious improvements. Seminar~\cite{seminar} adopts a simple multi-stage training algorithm to improve performance. It can be seen that compared with existing methods, our method can also achieve the best performance on the Cityscapes dataset. Specifically, AGMM++ obtains 65.2\% mIoU with 20 clicks, 70.1\% mIoU with 50 clicks, and 72.0\% mIoU with 100 clicks, respectively, outperforming the existing state-of-art methods by a large margin.

\begin{table}
	\setlength{\abovecaptionskip}{0.pt}
	\setlength{\belowcaptionskip}{-0.em}
	\centering
	\footnotesize
 \caption{Experimental results of the point-supervised WSSS methods on the Cityscapes val set. We use ResNet-101 as backbone.}
\begin{threeparttable}
	\begin{tabular}{l|cccc}
		\toprule[1.2pt]
		\textbf{Method} &\multicolumn{4}{c}{\textbf{Cityscapes}}\\
		\cline{2-5}
		&\textbf{20 clicks} &\textbf{50 clicks} &\textbf{100 clicks} &\textbf{full}\\
		\hline
		Baseline &53.5 &60.3 &64.2 &78.6\\
		\hline
		DenseCRF Loss~\cite{normcut} &54.2 &61.6 &65.5 &-\\
        Seminar~\cite{seminar} &57.1 &63.0 &66.1 &-\\
		TEL~\cite{TEL} &56.3 &62.8 &67.6 &-\\
		AGMM~\cite{AGMM} &62.1 &68.3 &71.6 &-\\
        \hline
        \rowcolor{mygray}
		AGMM++ &\textbf{65.2} &\textbf{70.1} &\textbf{72.0} &-\\
		\toprule[1.2pt]
	\end{tabular}
    \end{threeparttable}       
\label{table:city_point}
\end{table}

\begin{table}
	\setlength{\abovecaptionskip}{0.pt}
	\setlength{\belowcaptionskip}{-0.em}
	\centering
	\footnotesize
 \caption{Experimental results of the scribble-supervised WSSS methods on the Pascal validation set. Experimental settings with backbone, multi-stage training, and CRF post-processing are also considered.}
\begin{threeparttable}
	\begin{tabular}{l|c|c|c|c|c}
		\toprule[1.2pt]
		\textbf{Method}
		&\textbf{Pub.}
		&\textbf{Back.} &\textbf{Mul.}
		&\textbf{CRF} &\textbf{mIoU}\\
		\hline
		ScribbleSup~\cite{Scribblesup} &CVPR16 &R101 &$\checkmark$ &$\checkmark$ &63.1\\
		RAWKS~\cite{random-walk} &CVPR17 &R101 &$\checkmark$ &$\checkmark$ &61.4\\
		GraphNet~\cite{graphnet} &MM18 &R101 &$\checkmark$ &- &70.3\\
		NormCut~\cite{normcut} &CVPR18 &R101 &$\checkmark$ &- &72.8\\
		GridCRF~\cite{GridCut} &CVPR19 &R101 &$\checkmark$ &$\checkmark$ &72.8\\
		BPG~\cite{BPG} &IJCAI19 &R101 &- &- &73.2\\
        SPML~\cite{SPML} &ICLR21 &R101 &$\checkmark$ &- &74.2\\
		URSS~\cite{URSS} &ICCV21 &R101 &$\checkmark$ &- &74.6\\
		PSI~\cite{PSI} &ICCV21 &R101 &- &- &74.9\\
		A2GNN~\cite{A2GNN} &PAMI21 &R101 &$\checkmark$ &$\checkmark$ &74.3\\
        DBFNet~\cite{DBFNet} &TIP22 &R101 &- &- &72.5\\
		TEL~\cite{TEL} &CVPR22 &R101 &- &- &75.8\\
        TSCD+SEAM~\cite{TSCD} &AAAI23 &MiT-B1 &$\checkmark$ &$\checkmark$ &76.8\\
        CDL~\cite{CDL} &IJCV23 &R101 &$\checkmark$ &$\checkmark$ &76.1\\
		AGMM~\cite{AGMM} &CVPR23 &R101 &- &- &76.4\\
        \hline
        \rowcolor{mygray}
        AGMM++ &- &R101 &- &- &77.5\\
        \rowcolor{mygray}
        AGMM++ &- &R101 &- &$\checkmark$ &78.3\\
        \rowcolor{mygray}
        AGMM++ &- &ViT-B &- &- &78.7\\
        \rowcolor{mygray}
        AGMM++ &- &ViT-B &- &$\checkmark$ &\textbf{80.3}\\
		\toprule[1.2pt]
	\end{tabular}
    \end{threeparttable}
\label{pascal_scri}
\end{table}%

\subsubsection{Scribble-Supervised WSSS}

The results of scribble-supervised WSSS on the PASCAL dataset are shown in Table~\ref{pascal_scri}. It can be seen that most existing methods employ CRF~\cite{deeplab} during testing, which can bring about 3\% mIoU improvements. Multi-stage training strategy is also widely employed, which requires time-consuming training. It is worth noting that RAWKS~\cite{random-walk}, BPG~\cite{BPG}, and SPML~\cite{SPML} create extra edge information \cite{edge} for supervision. In our experiments, we first discard these settings to evaluate the pure effectiveness of our proposed AGMM++. Our AGMM++ achieves 77.5\% mIoU and outperforms all existing methods. \zhun{Furthermore, we demonstrate that our AGMM++ framework can achieve additional improvements by incorporating CRF. Moreover, by employing ViT-B as the backbone, our AGMM++ framework achieves an impressive mIoU of 80.3\%, underscoring the versatility and effectiveness of our method.}

\begin{table}
	\setlength{\abovecaptionskip}{0.pt}
	\setlength{\belowcaptionskip}{-0.em}
	\centering
	\footnotesize
 \caption{Experimental results of the box-supervised WSSS methods on the Pascal validation set.}
\begin{threeparttable}
	\begin{tabular}{l|c|c|c}
		\toprule[1.2pt]
		\textbf{Method}
		&\textbf{Publications}
		&\textbf{Backbone} &\textbf{mIoU}\\
		\hline
		Box-Sup~\cite{boxsup} &CVPR15 &VGG16 &62.0\\
		WSSL~\cite{WSSL} &ICCV15 &VGG16 &60.6\\
		SDI~\cite{SDI} &CVPR17 &R101 &69.4\\
		GraphNet~\cite{graphnet} &MM18 &R101 &65.6\\
		BCM~\cite{BCM} &CVPR19 &R101 &70.2\\
		SEAM~\cite{seam} &CVPR20 &R38 &70.6\\
        A2GNN~\cite{A2GNN} &PAMI21 &R101 &73.8\\
        SPML~\cite{SPML} &ICLR21 &R101 &73.5\\
		BBAM~\cite{bbam} &CVPR21 &R101 &73.7\\
		BAP~\cite{BAP} &CVPR21 &R101 &72.4\\
		TEL+SEAM~\cite{TEL} &CVPR22 &R101 &74.1\\
        TSCD+SEAM~\cite{TSCD} &AAAI23 &MiT-B1 &74.0\\
        CDL~\cite{CDL} &IJCV23 &R101 &74.4\\
        \hline
        \rowcolor{mygray}
        AGMM++ &- &R101 &75.5\\
        \rowcolor{mygray}
        AGMM++ &- &ViT-B &\textbf{77.6}\\
		\toprule[1.2pt]
	\end{tabular}
    \end{threeparttable}
\label{pascal_box}
\end{table}%

\begin{table}
	\setlength{\abovecaptionskip}{0.pt}
	\setlength{\belowcaptionskip}{-0.em}
	\centering
	\footnotesize
 \caption{Experimental results of the block-supervised WSSS methods on the ADE20K val set. We use ResNet-101 as backbone.}
\begin{threeparttable}
	\begin{tabular}{l|cccc}
		\toprule[1.2pt]
		\textbf{Method} &\multicolumn{4}{c}{\textbf{ADE20K}}\\
		\cline{2-5}
		&\textbf{10\%} &\textbf{20\%} &\textbf{50\%} &\textbf{full}\\
		\hline
		Baseline &30.8 &33.1 &37.2 &42.5\\
		\hline
		DenseCRF Loss~\cite{normcut} &31.2 &34.0 &37.4 &-\\
        Seminar~\cite{seminar} &34.1 &35.5 &38.8 &-\\
		TEL~\cite{TEL} &34.3 &36.0 &39.2 &-\\
		AGMM~\cite{AGMM} &36.0 &37.2 &39.8 &-\\
        \hline
        \rowcolor{mygray}
		AGMM++ &\textbf{37.1} &\textbf{37.4} &\textbf{40.0} &-\\
		\toprule[1.2pt]
	\end{tabular}
    \end{threeparttable}       
\label{table:ade_block}
\end{table}

\subsubsection{Box-Supervised WSSS}

We conduct box-supervised WSSS on the PASCAL VOC dataset, as shown in Table~\ref{pascal_box}. We follow the settings of A2GNN~\cite{A2GNN}, which first generate initial seed labels from bounding boxes using SEAM~\cite{seam}. Then, Grab-cut labels are generated by SDI~\cite{SDI}. Multi-scale testing and CRF~\cite{deeplab} post-processing are adopted by all methods in Table~\ref{pascal_box}. We further evaluate two recent methods~\cite{TEL, TSCD} on box-supervised WSSS experiments. It can be seen that our AGMM++ can also gain the best performance. Specifically, with ViT-B~\cite{vit} as the backbone, AGMM++ can achieve the highest 76.7\% mIoU.

\subsubsection{Block-Supervised WSSS}
As shown in Table~\ref{table:ade_block}, we conduct block-supervised WSSS on the ADE20K dataset. Following the previous work TEL~\cite{TEL}, DeeplabV3++~\cite{deeplabv3} and ResNet-101~\cite{resnet} are used for all the experiments. Compared with existing state-of-the-art methods~\cite{normcut, seminar, TEL, AGMM}, our AGMM++ also achieves the best performance under different ratios of block annotations. Specifically, AGMM++ achieves 37.1\% mIoU with 10\% labels, 37.4\% mIoU with 20\% labels, and 40.0\% mIoU with 50\% labels, respectively. 

\subsection{Ablation Studies}

In this section, we conduct extensive ablation experiments to evaluate the effectiveness of our proposed AGMM++. First, we present the results of GMM predictions and demonstrate the improvements of pseudo labels quality. Second, we compare our improved AGMM++ with our previous version AGMM~\cite{AGMM}. Third, we conduct extensive experiments to evaluate the hyper-parameters settings and loss functions in the proposed AGMM++. Finally, we compare our AGMM++ with the non-adaptive baselines to verify the importance of the adaptive manner. 

\begin{figure}
	\centering
	\includegraphics[width=1\linewidth]{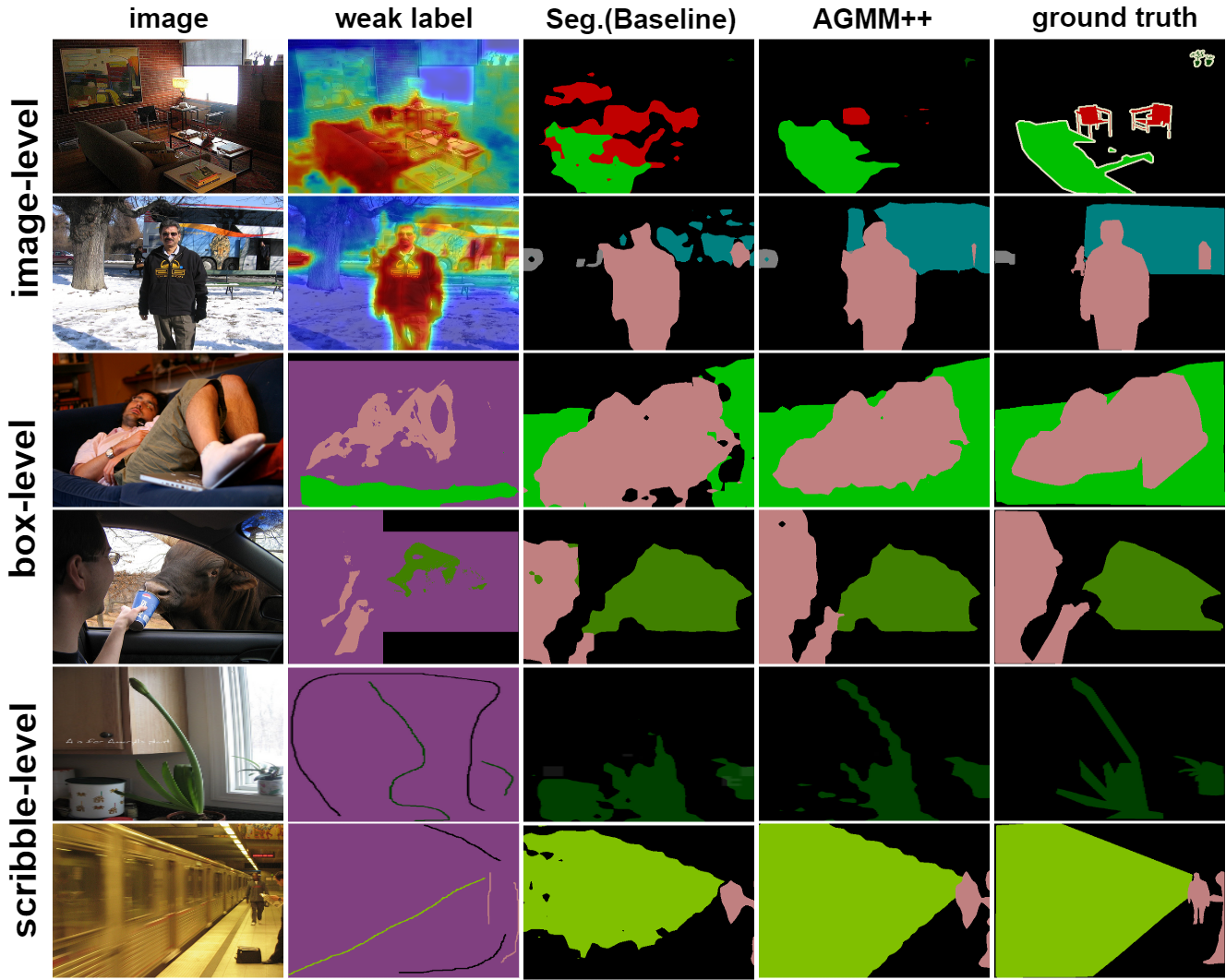}
	\caption{The visualization results of the generated pseudo labels based on initial weak labels (CAMs, boxes, and scribbles). }
	\label{voc_label_update}
\end{figure}

\subsubsection{Quality of GMM Predictions}\label{sec4_2_1}

\begin{table}
	\setlength{\abovecaptionskip}{0.pt}
	\setlength{\belowcaptionskip}{-0.em}
	\centering
	\footnotesize
 \caption{Ablation study for pseudo labels generations. The results are reported on the PASCAL dataset with ViT-B as backbone (without CRF).}
\begin{threeparttable}
	\begin{tabular}{l|cccc|c|c}
		\toprule[1.2pt]
		\textbf{Method} &\bm{$L_{self}$} &\bm{$L_{weak}$} &\bm{$L_{con}$} &\textbf{OEM} &\textbf{pseudo} &\textbf{seg.}\\
		\hline
		ToCo~\cite{toco} &-	&- &- &- &70.5 &68.1\\
		\hline
		&$\checkmark$ &- &- &- &71.5 &69.1\\
		AGMM++ &$\checkmark$ &$\checkmark$ &- &- &72.1 &69.5\\
        &$\checkmark$ &$\checkmark$ &$\checkmark$ &- &72.5 &70.2\\
		&$\checkmark$ &$\checkmark$	&$\checkmark$ &$\checkmark$ &\textbf{73.3} &\textbf{70.8}\\
		\toprule[1.2pt]
	\end{tabular}
    \end{threeparttable}        
\label{abl_pseudo}
\end{table}%

We first present the results of generating pseudo labels based \zhun{different} initial weak labels. Visualization results are shown in Fig.~\ref{voc_label_update}. Specifically, we compare our GMM predictions with initial weak labels, \emph{i.e.}, CAMs, boxes, and scribbles labels. Previous methods~\cite{seminar, RIB, seam, A2GNN} mainly generate pseudo labels through the segmentation branch (Seg. in Fig.~\ref{voc_label_update}). In our AGMM++, we use the GMM to generate GMM predictions as pseudo labels. It can be seen that compared with the partial weak labels and the coarse pseudo labels, our AGMM++ can generate better pseudo labels and provide more reliable and complete supervision.  

\begin{figure}
	\centering
	\includegraphics[width=1\linewidth]{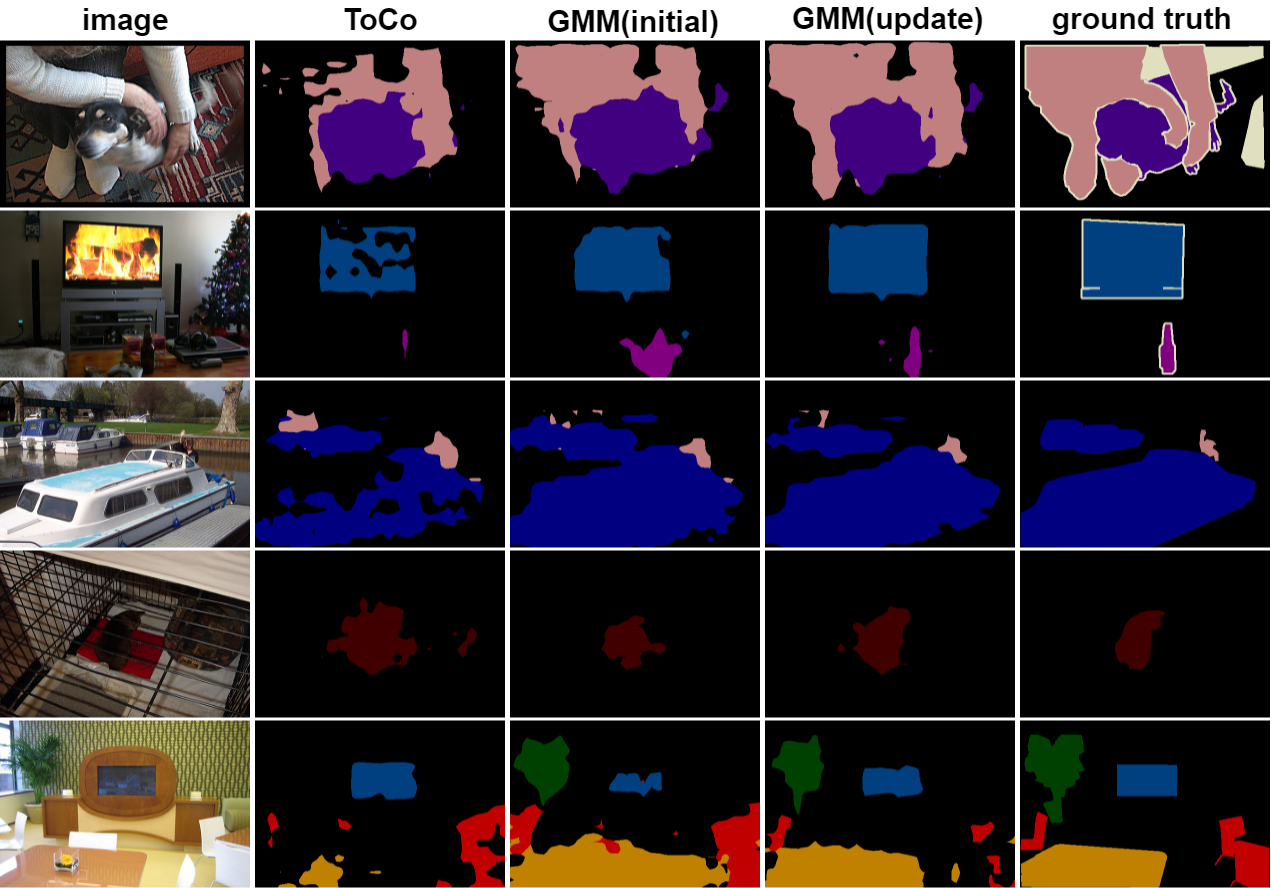}
	\caption{Qualitative results of pseudo labels on the PASCAL training set under image-level supervised settings. }
	\label{voc_pseudo}
\end{figure}

\zhun{We then investigate the quality of pseudo labels image-level supervised WSSS on PASCAL dataset. As shown in Table~\ref{abl_pseudo}, the initial CAMs generated by ToCo~\cite{toco} can achieve 70.5\% mIoU. Based on the ToCo~\cite{toco} CAMs, our AGMM++ generates GMM predictions as pseudo labels. $L_{self}$, $L_{weak}$, $L_{con}$, and OEM can consistently improve the quality of pseudo labels. We also compare our approach with other methods in Table~\ref{pseudo}. We can find that the pseudo labels generated by our AGMM++ can achieve 73.3\% mIoU, which significantly outperforms previous methods. We further present some visualization results of our GMM predictions in Fig.~\ref{voc_pseudo}. Clearly, our GMM can generate better pseudo labels by employing cross-label constraints on the initial CAMs labels and the pseudo labels can be further improved by using the OEM algorithm.}


\begin{table}
	\setlength{\abovecaptionskip}{0.pt}
	\setlength{\belowcaptionskip}{-0.em}
	\centering
	\footnotesize
 \caption{Comparison results of the pseudo labels with image-level labels on the PASCAL dataset.}
\begin{threeparttable}
	\begin{tabular}{l|c|c|c}
		\toprule[1.2pt]
		\textbf{Method}
		&\textbf{Publications}
		&\textbf{Backbone} &\textbf{Pseudo label}\\
        \hline
        1stage~\cite{singlestage} &CVPR22 &R38 &66.9\\
		AFA~\cite{AFA} &CVPR22 &MiT-B1 &68.7\\
        SLRNet~\cite{SLRNet} &IJCV22 &R38 &67.1\\
        ViT-PCM~\cite{Vit-pcm} &ECCV22 &ViT-B &67.7\\
        TSCD~\cite{TSCD} &AAAI23 &MiT-B1 &69.0\\
        ToCo~\cite{toco} &CVPR23 &ViT-B &70.5\\
        \hline
        \rowcolor{mygray}
        AGMM++ &- &ViT-B &\textbf{73.3}\\
		\toprule[1.2pt]
	\end{tabular}
    \end{threeparttable}
\label{pseudo}
\end{table}%

\begin{figure}
	\centering
	\includegraphics[width=1\linewidth]{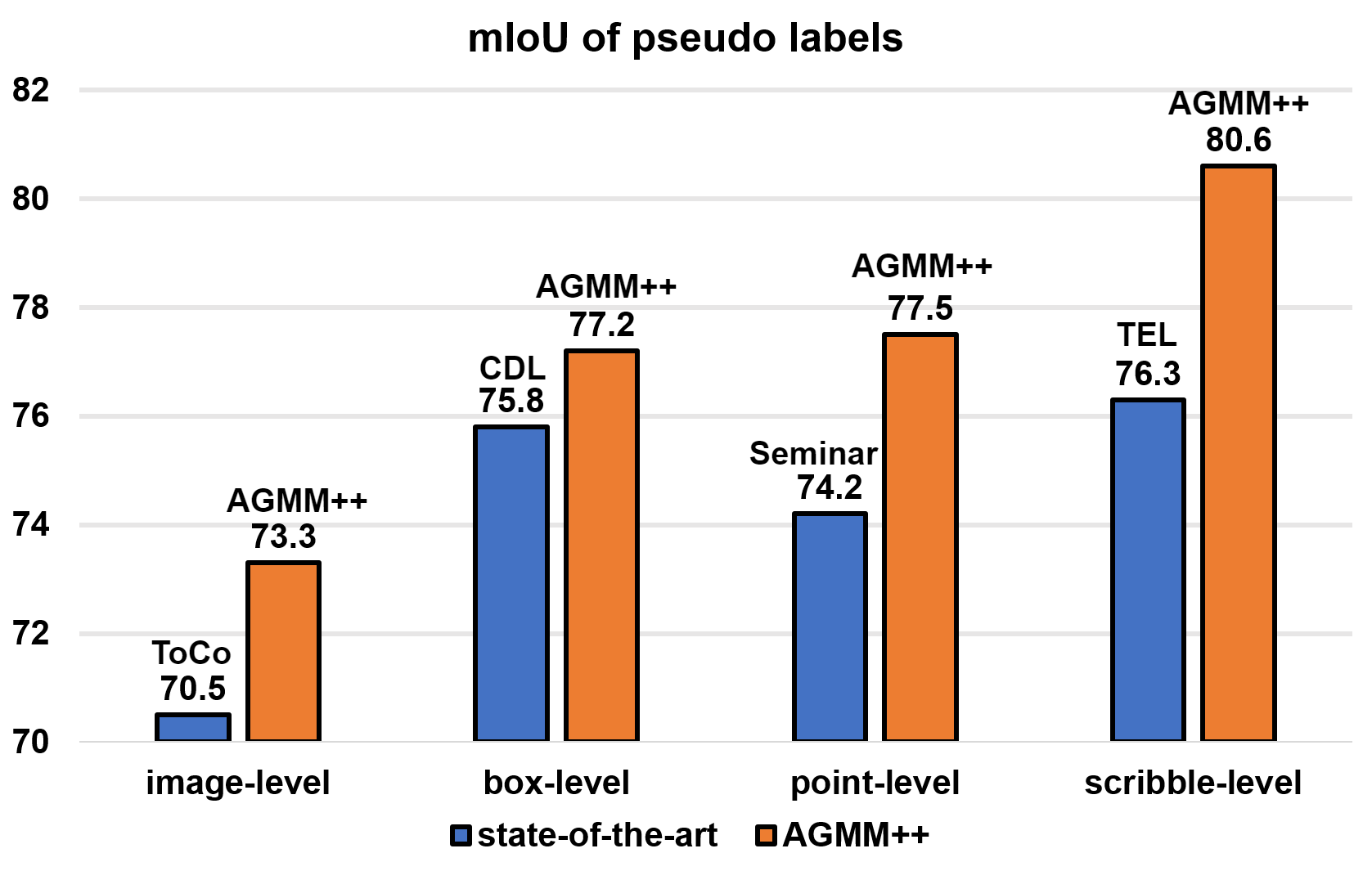}
	\caption{Comparisons with different state-of-the-art methods on the PASCAL training set under different supervised settings. }
	\label{voc_compare}
\end{figure}

\begin{figure}
	\centering
	\includegraphics[width=1\linewidth]{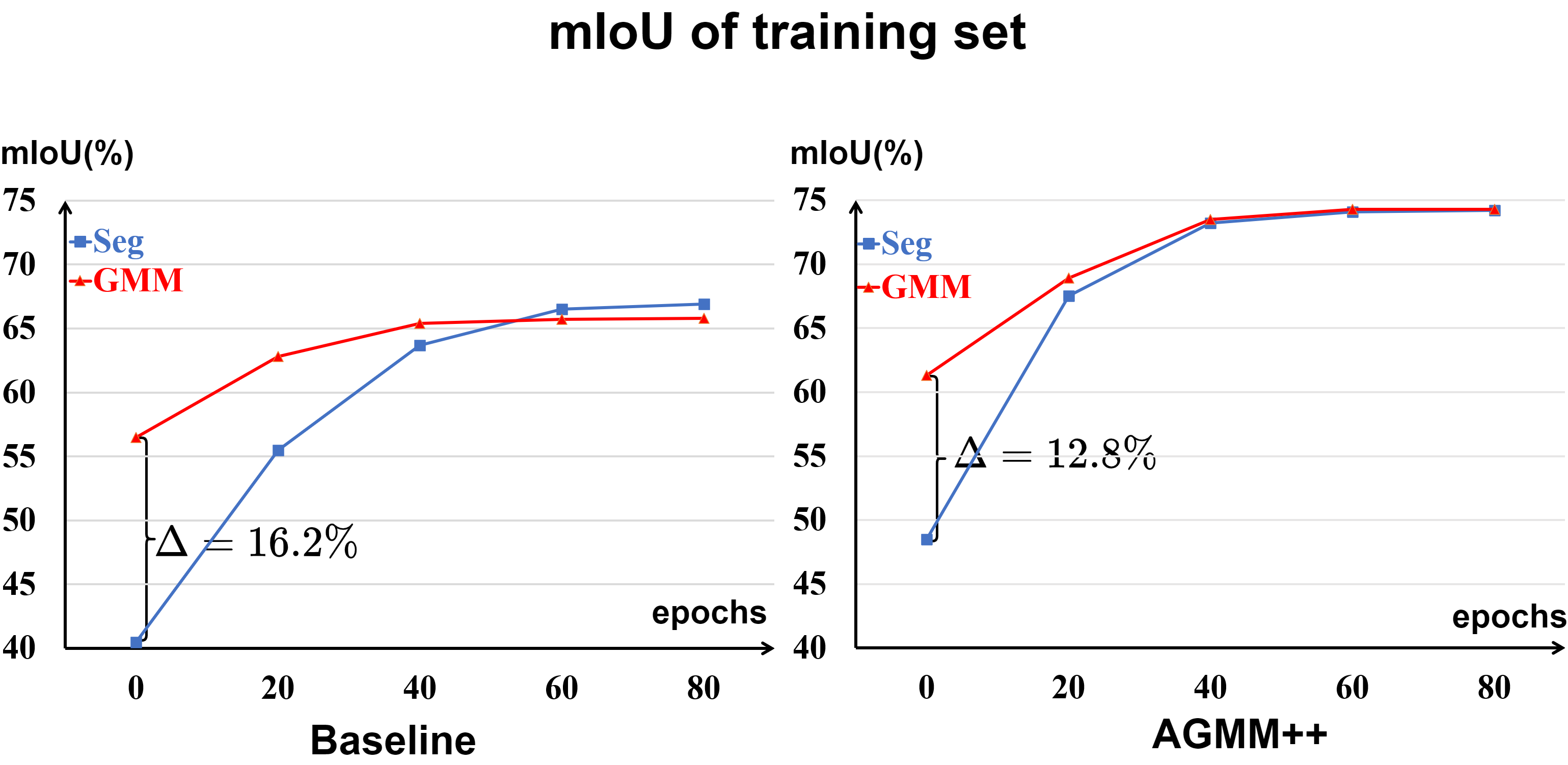}
	\caption{Comparison between segmentation and GMM predictions during training. ${\Delta}$ denotes the difference of mIoU between segmentation and GMM predictions.}
	\label{line}
\end{figure}

\begin{table}
	\setlength{\abovecaptionskip}{0.pt}
	\setlength{\belowcaptionskip}{-0.em}
	\centering
	\footnotesize
 \caption{Comparison with AGMM~\cite{AGMM} on the PASCAL dataset under point- and scribble-supervised settings.}
\begin{threeparttable}
	\begin{tabular}{l|ccc|cc}
		\toprule[1.2pt]
		\textbf{Method}
		&\textbf{OEM} &\bm{$L_{con}$}(old)
		&\bm{$L_{con}$} &\textbf{point} &\textbf{scribble}\\
        \hline
        AGMM~\cite{AGMM} &- &$\checkmark$ &- &69.6 &76.4\\
        \hline
		AGMM++ &$\checkmark$ &$\checkmark$ &- &71.2 &77.0\\
        \rowcolor{mygray}
        AGMM++ &$\checkmark$ &- &$\checkmark$ &\textbf{72.3} &\textbf{77.5}\\
		\toprule[1.2pt]
	\end{tabular}
    \end{threeparttable}
\label{compare_agmm}
\end{table}%

\begin{table}
	\setlength{\abovecaptionskip}{0.pt}
	\setlength{\belowcaptionskip}{-0.em}
	\centering
	\footnotesize
 \caption{Efficiency comparison with AGMM~\cite{AGMM} on the PASCAL dataset under point- and scribble-supervised settings. We use ViT-B as the backbone and a $448\times448$ size image for evaluation. Baseline means without GMM formulation.}
\begin{threeparttable}
	\begin{tabular}{l|c|cc}
		\toprule[1.2pt]
		\textbf{Method}
		&\textbf{FLOPs(G)} &\textbf{point} &\textbf{scribble}\\
        \hline
        Baseline &83.2 &59.2 &67.3\\
        \hline
		AGMM~\cite{AGMM} &117.5 &71.4 &77.0\\
        \rowcolor{mygray}
        AGMM++ &96.3 &\textbf{75.6} &\textbf{78.7}\\
		\toprule[1.2pt]
	\end{tabular}
    \end{threeparttable}
\label{efficiency}
\end{table}%


In addition, we provide the comparisons with different existing state-of-the-art methods under different supervised settings in Fig.~\ref{voc_compare}. The results are reported on PASCAL training set. 
\zhun{It is evident that our proposed AGMM++ method consistently generates higher-quality pseudo labels compared to various state-of-the-art methods across different supervised settings.}

We further analyze the GMM predictions during the training process. In Fig.~\ref{line}, we report the results on the PASCAL dataset under point-supervised settings. The baseline method represents using partial cross-entropy loss $L_{seg}$ for supervision only. It can be seen that at the beginning of training, the GMM predictions gain a higher accuracy than the segmentation predictions by a large margin. In our AGMM++, we utilize the $L_{self}$ to constrain the consistency between segmentation and GMM predictions, leading to better supervision for the segmentation model.

\begin{table}
	\setlength{\abovecaptionskip}{0.pt}
	\setlength{\belowcaptionskip}{-0.em}
	\centering
	\footnotesize
 \caption{Ablation study for the PASCAL dataset under different settings. We use ViT-B as the backbone (without CRF).}
\begin{threeparttable}
	\begin{tabular}{ccccc|ccc}
		\toprule[1.2pt]
		\bm{$L_{seg}$} &\bm{$L_{self}$} &\bm{$L_{weak}$} &\bm{$L_{con}$} &\textbf{OEM} &\textbf{box} &\textbf{point} &\textbf{scrib.}\\
		\hline
		$\checkmark$ &-	&- &- &- &65.4 &63.2 &69.0\\
        \hline
		$\checkmark$ &$\checkmark$	&-	&- &- &72.3 &72.0 &73.2\\
		$\checkmark$ &$\checkmark$	&$\checkmark$	&-	&- &73.2 &74.5 &76.1\\
        $\checkmark$ &$\checkmark$	&$\checkmark$ &$\checkmark$	&- &74.5 &75.0 &77.2\\
        \rowcolor{mygray}
		$\checkmark$ &$\checkmark$	&$\checkmark$	&$\checkmark$ &$\checkmark$	&\textbf{75.5} &\textbf{75.6} &\textbf{78.7}\\
		\toprule[1.2pt]
	\end{tabular}
    \end{threeparttable}       
\label{table:abl_loss_voc}
\end{table}%

\begin{table}
	\setlength{\abovecaptionskip}{0.pt}
	\setlength{\belowcaptionskip}{-0.em}
	\centering
	\footnotesize
 \caption{Ablation study for Cityscapes and ADE20K datasets. We use ResNet as the backbone. 20 clicks for Cityscapes dataset and 10\% blocks for ADE20K dataset.}
\begin{threeparttable}
	\begin{tabular}{ccccc|cc}
		\toprule[1.2pt]
		\bm{$L_{seg}$} &\bm{$L_{self}$} &\bm{$L_{weak}$} &\bm{$L_{con}$} &\textbf{OEM} &\textbf{City.(20)} &\textbf{ADE.(10\%)}\\
		\hline
		$\checkmark$ &-	&-	&-  &-&53.5 &30.8\\
		\hline
		$\checkmark$ &$\checkmark$	&-	&- &- &60.5 &34.2\\
		$\checkmark$ &$\checkmark$	&$\checkmark$	&-	&- &63.5 &35.6\\
        $\checkmark$ &$\checkmark$	&$\checkmark$	&$\checkmark$	&- &64.8 &36.7\\
        \rowcolor{mygray}
		$\checkmark$ &$\checkmark$	&$\checkmark$	&$\checkmark$	&$\checkmark$ &\textbf{65.2} &\textbf{37.1}\\
		\toprule[1.2pt]
	\end{tabular}
    \end{threeparttable}       
\label{table:abl_loss_other}
\end{table}%

\begin{table}
	\setlength{\abovecaptionskip}{0.pt}
	\setlength{\belowcaptionskip}{-0.em}
	\centering
	\footnotesize
 \caption{Evaluation of $\lambda_{G}$ in on PASCAL dataset. We report the mIoU results with point- and scribble-supervised settings. $\lambda_{seg}$ is set to 1.}
\begin{threeparttable}
	\begin{tabular}{c|cccccc}
		\toprule[1.2pt]
		\bm{$\lambda_{G}$} &\textbf{0.1} &\textbf{0.2} &\textbf{0.5} &\textbf{0.8} &\textbf{1.0} &\textbf{1.5}\\
		\hline
		point-supervised &68.5 &70.1 &70.6 &71.2 &\textbf{72.3} &71.3\\
		scribble-supervised &76.0 &76.8 &77.9 &78.1 &\textbf{78.3} &78.2\\
		\toprule[1.2pt]
	\end{tabular}
    \end{threeparttable}        
\label{lambda}
\end{table}%

\begin{table}
	\setlength{\abovecaptionskip}{0.pt}
	\setlength{\belowcaptionskip}{-0.em}
	\centering
	\footnotesize
 \caption{Evaluation of the adaptive manner in AGMM++. Results reported on the PASCAL dataset. Hard and soft means one-hot and probabilistic labels, respectively. Online means our AGMM++ generates pseudo labels online. SG means stop-gradients of GMM.}
\begin{threeparttable}
	\begin{tabular}{l|ccc|cc}
		\toprule[1.2pt]
		\textbf{Method} &\textbf{hard} &\textbf{soft} &\textbf{online} &\textbf{point} &\textbf{scrib.}\\
		\hline
		Baseline &- &- &- &59.2 &67.3\\
		Baseline (+MT) &$\checkmark$ &- &- &66.3 &72.4\\
		\hline
		Label Assignment
		&$\checkmark$ &- &$\checkmark$ &66.5 &73.4\\
		AGMM++ (SG) &- &$\checkmark$ &$\checkmark$ &67.4 &74.6\\
        \hline
        \rowcolor{mygray}
		AGMM++ &- &$\checkmark$ &$\checkmark$ &\textbf{72.3} &\textbf{78.3}\\
		\toprule[1.2pt]
	\end{tabular}
    \end{threeparttable}
\label{adapt_compare}
\end{table}%

\subsubsection{Comparison with AGMM}

Compared with our previous version AGMM~\cite{AGMM}, in our improved AGMM++, we further develop a novel OEM algorithm to optimize the GMM with a novel maximization loss $L_{GMM}$. In addition, we re-define the contrastive loss $L_{con}$ to provide stronger supervision for each pixel. To evaluate the effectiveness of these two modifications, we conduct ablation experiments on the PASCAL dataset under point- and scribble-supervised settings. As shown in Table~\ref{compare_agmm}, compared with our previous AGMM~\cite{AGMM}, the OEM and the re-defined $L_{con}$ can consistently improve the performance, especially under the point-supervised setting. This validates the effectiveness of our modifications in AGMM++.

We further evaluate the model efficiency of our AGMM++ in Table~\ref{efficiency}. In our improved AGMM++, we introduce OEM to refine the GMM predictions, which increases the computation costs. Thus, we add a convolutional layer to squeeze the feature channels in the GMM module and largely reduce the computation costs, as described in Section~\ref{sec3_2}. As shown in Table~\ref{efficiency}, AGMM++ is more efficient than the previous version and achieves better performance.

\subsubsection{Analysis of Loss Functions and Hyper-parameters}\label{sec4_loss}

\zhun{The evaluation of loss functions are presented in Table~\ref{abl_pseudo}, Table~\ref{table:abl_loss_voc} and Table~\ref{table:abl_loss_other}. Specifically, Table~\ref{abl_pseudo} shows the results on image-level setting, Table~\ref{table:abl_loss_voc} reports the results on other supervised settings on PASCAL dataset, and Table~\ref{table:abl_loss_other} presents the results of points and block labels on Cityscapes and ADE20k datasets. Through these extensive experiments, we observe that AGMM++ consistently improves the performance of the baseline, and all the proposed components consistently enhance the performance.}




We then investigate the balance between $\lambda_{G}$ and $\lambda_{seg}$ under the points and scribble settings. 
For clear comparisons, we do not use CRF post-processing in this experiment. As shown in Table~\ref{lambda}, $\lambda_{G}=1$ yields the best performance, indicating that the proposed $L_{GMM}$ is as important as the segmentation loss $L_{seg}$.

\begin{figure}
	\centering
	\includegraphics[width=1\linewidth]{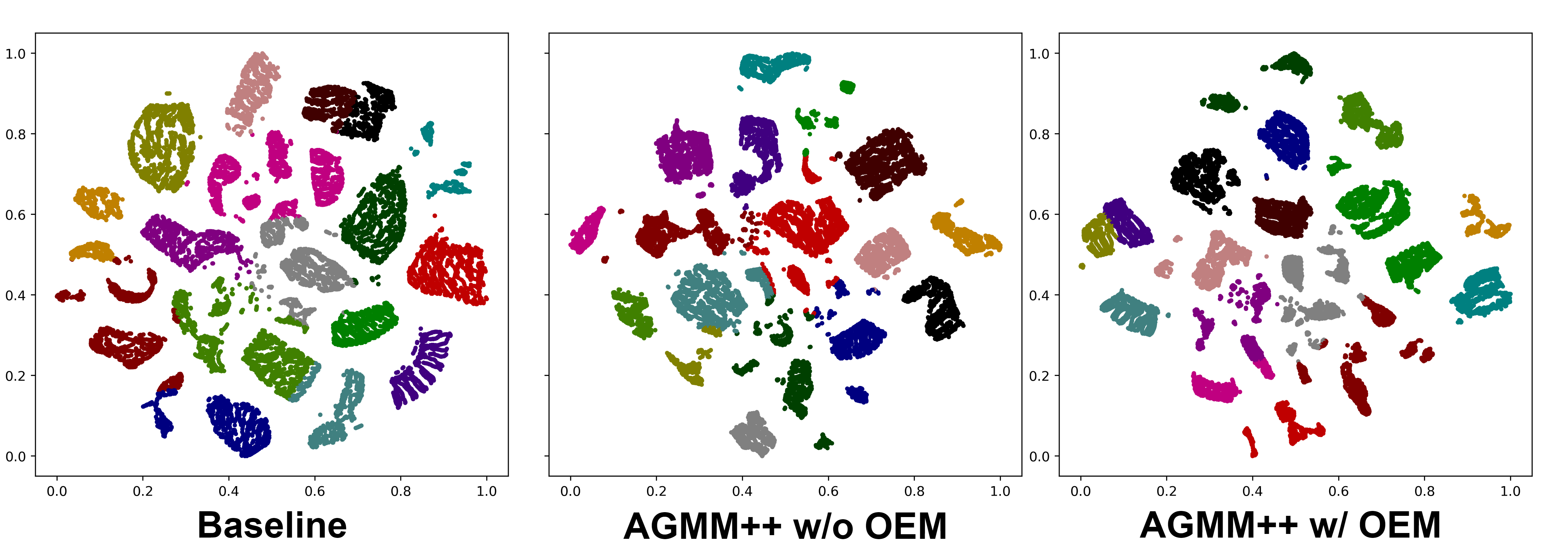}
	\caption{Visualization of the high-level feature space learned by baseline, AGMM++ (without OEM), and AGMM++ (with OEM), respectively.}
	\label{tsne}
\end{figure}

\begin{figure*}
	\centering
	\includegraphics[width=0.95\linewidth]{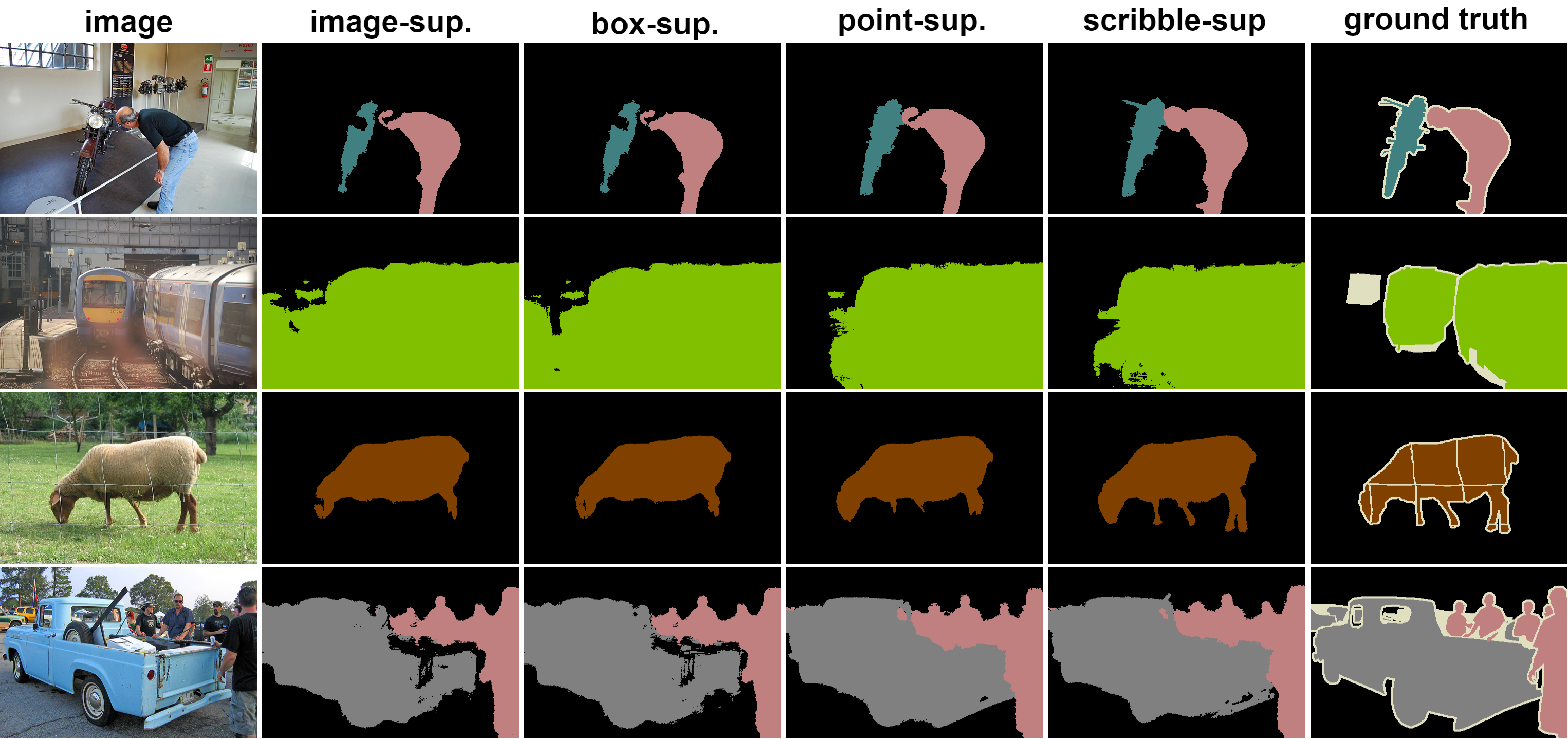}
	\caption{Qualitative segmentation results for the Pascal dataset under different supervised settings. From left to right, the segmentation results are generated from AGMM++ supervised by image-, box-, point-, and scribble-level labels.}
	\label{pascal_val}
\end{figure*}

\begin{figure*}
	\centering
	\includegraphics[width=0.95\linewidth]{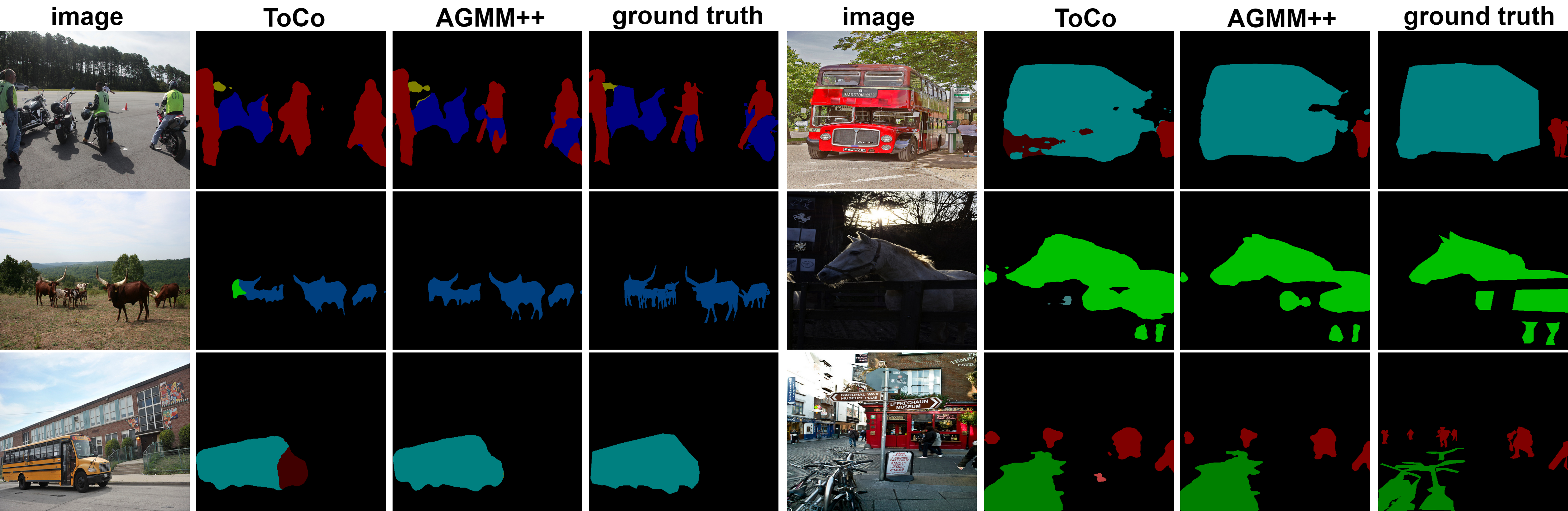}
	\caption{Qualitative segmentation results for the COCO dataset supervised by image-level labels. To evaluate the effectiveness, we compare our results with the results of ToCo~\cite{toco}.}
	\label{coco_val}
\end{figure*}

\begin{figure*}
	\centering
	\includegraphics[width=0.95\linewidth]{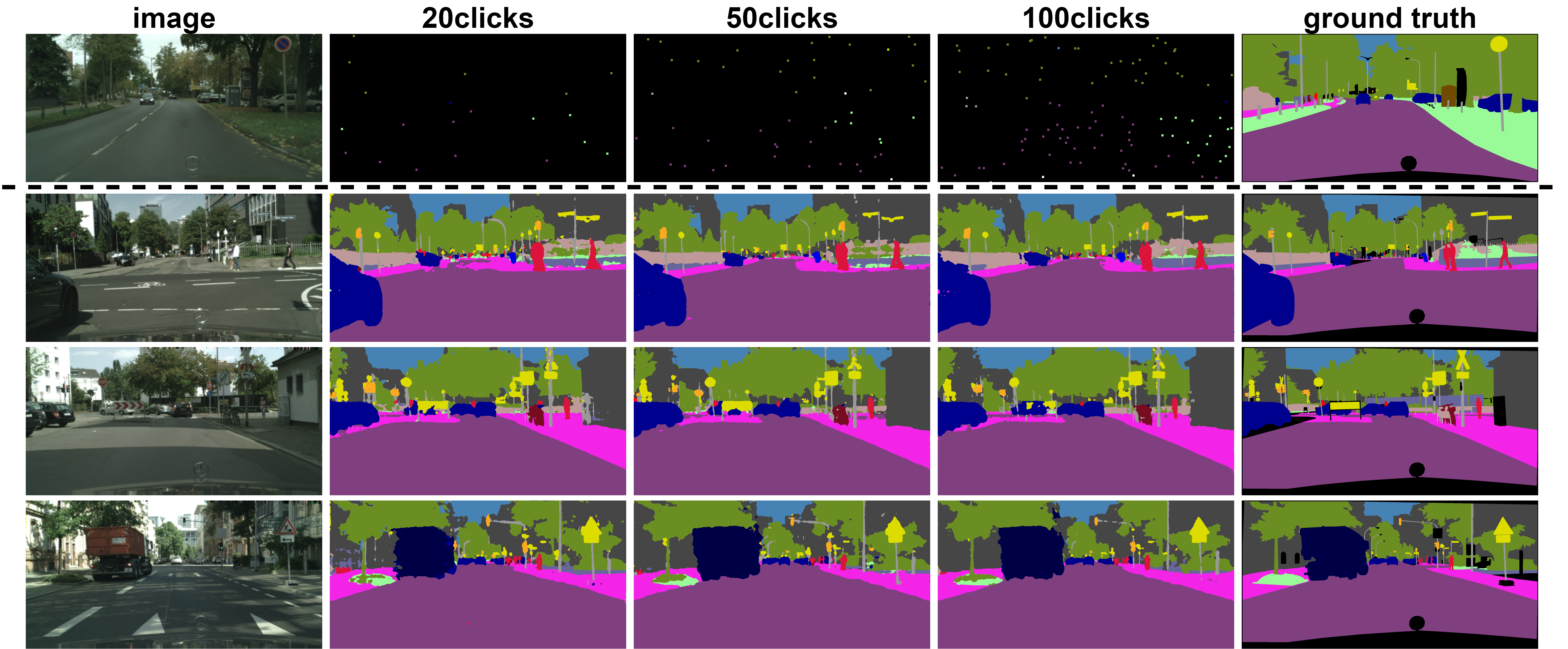}
	\caption{Qualitative segmentation results of Cityscapes under point-supervised settings. The first row presents samples of point annotations.}
	\label{city_val}
\end{figure*}

\begin{figure}
	\centering
	\includegraphics[width=0.95\linewidth]{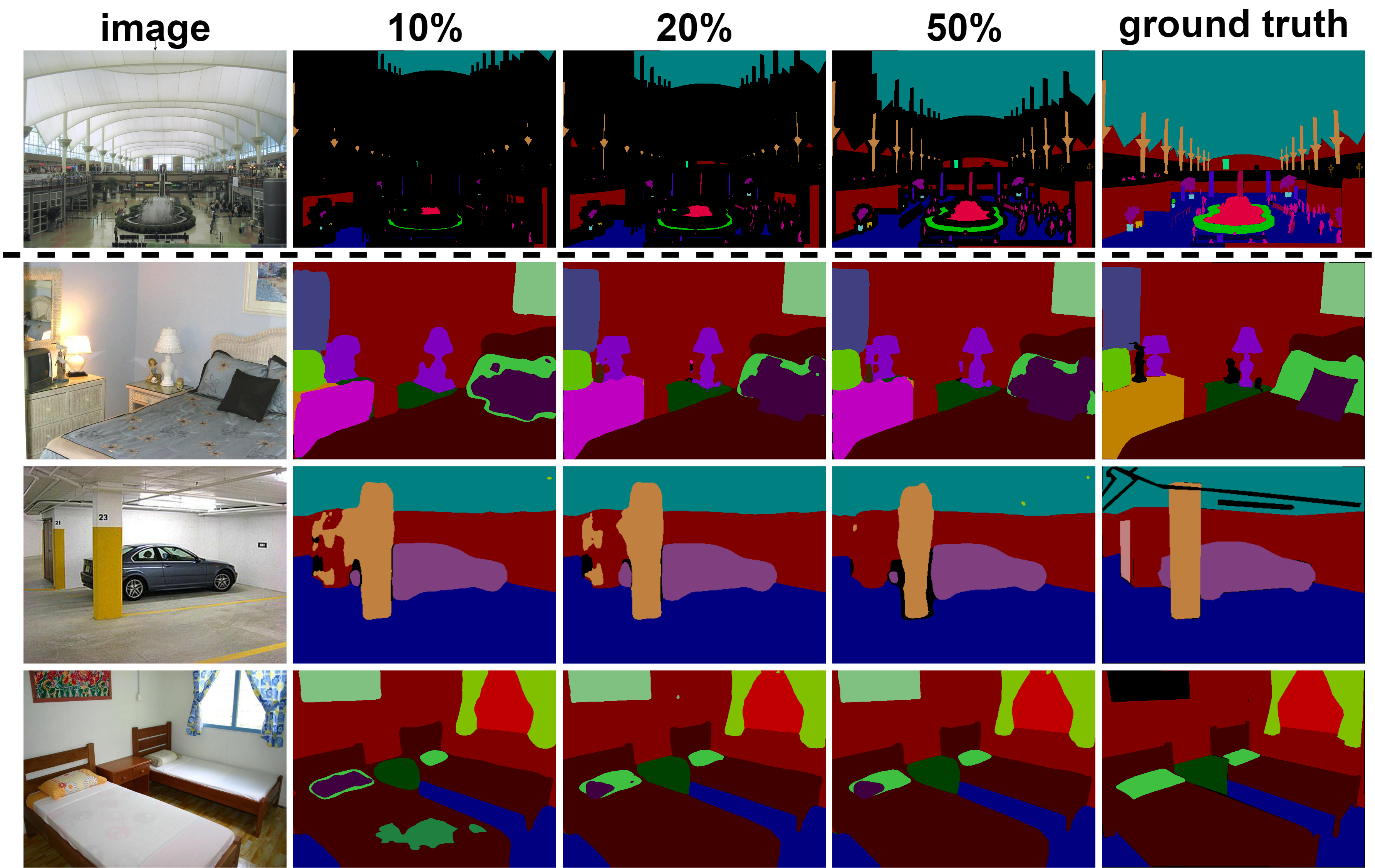}
	\caption{Qualitative segmentation results for the ADE20K dataset under block-supervised settings. The first row presents samples of block annotations, which are provided by TEL~\cite{TEL}.}
	\label{ade_val}
\end{figure}

\subsubsection{Comparison with Non-adaptive Baseline}

To prove our core insight, \emph{i.e.}, modeling the label distributions with adaptive GMM, we further compare our proposed AGMM++ with several non-adaptive baselines. 
\zhun{
\begin{itemize}[leftmargin=.25in]
    \item \textbf{Baseline:} This approach is implemented by using $L_{seg}$ for supervision. We also evaluate the effectiveness of Multi-stage Training (MT) for baseline.
    
    \item \textbf{Label Assignment:} Instead of GMM formulation, we use a similarity-based label assignment method to generate hard one-hot pseudo labels for supervision. Specifically, we simply set a fixed threshold $d<{\sigma}$ to assign the unlabeled pixels to specific categories, \emph{i.e.}, an unlabeled pixel is assigned to $i_{th}$ category when:
\begin{equation}\label{condition}
	d<{\sigma}_{i}, d>{\sigma}_{j}, \forall{i, j}\in{K}, {j}\neq{i}.
\end{equation}If an unlabeled pixel is not satisfied with Eq.~\ref{condition}, we ignore this pixel during training.

    \item \textbf{AGMM++ (SG):} We stop the gradients of GMM optimization in our AGMM++. In this case, $L_{weak}$ is discarded and only the segmentation branch is optimized during training. 
\end{itemize}

We show the comparisons under the point and scribble WSSS settings in Table~\ref{adapt_compare}. The results show that it is not appropriate to roughly assign hard one-hot pseudo labels to the unlabeled pixels according to Eq.~\ref{condition} (refer to the row of Label Assignment), since we cannot set accurate thresholds for the similarity. In addition, stopping the gradients of GMM predictions leads to lower results, indicating the significance of optimizing GMM in our framework.
}

\subsection{Visualization Results}

\textbf{{Feature Distribution Analysis}.} To better understand our AGMM++, we provide an illustration of visualization of the high-dimension feature space using t-SNE~\cite{t-sne}. We show the results of the baseline method ToCo~\cite{toco}, AGMM++ (without OEM), and AGMM++ (with OEM) on image-level supervised WSSS on the PASCAL dataset. As shown in Fig.~\ref{tsne}, the decision boundaries of features generated by the baseline method are quite confusing, while our AGMM++ can learn more discriminative features. Specifically, with the back-propagation optimization of the proposed loss functions stated in Section~\ref{sec3_4}, the GMM pushes each pixel into specific Gaussian mixtures by leveraging the semantic correlation of different labels. Furthermore, the OEM algorithm demonstrates further performance improvement. These results provide an explanation for the effectiveness of AGMM++ from a feature-aware perspective.


\textbf{Qualitative Semantic Segmentation Results}. \zhun{Finally, various qualitative semantic segmentation results are shown in Figs.~\ref{pascal_val}, \ref{coco_val}, \ref{city_val}, and \ref{ade_val}. These visualizations reveal that our proposed method produces predictions with more consistent regions, smoother internal structures, and finer boundaries across all four datasets under different settings, surpassing the performance of other methods. This further validates the effectiveness of our approach. Additional qualitative results can be found in the supplementary material.}


\section{Conclusion}

In this paper, we propose a unified WSSS framework AGMM++, which is capable of solving different forms of weak labels (image-level labels, points, scribbles, boxes, and blocks). Our AGMM++ explores the inherent semantic relation among different pixels and employs cross-label constraints to improve the quality of pseudo labels. \zhun{AGMM++ models the label distributions by GMM and is progressively optimized during training, which can effectively improve the quality of pseudo labels for more robust supervision. Experiments conducted on four different datasets (PASCAL, COCO, Cityscapes, and ADE20K) with five kinds of weak labels show that our proposed AGMM++ achieves state-of-the-art results on all WSSS settings.

In our future work, we will investigate the applicability of our AGMM++ framework to refine pseudo labels in other tasks such as domain adaptive and semi-supervised semantic segmentation. Additionally, we aim to explore the potential of AGMM++ in handling datasets with highly imbalanced distributions. Furthermore, we believe that alternative probabilistic models, such as Bayes models~\cite{bayes}, hold promise for modeling label distributions, and we intend to explore their use in future research.}

\bibliography{ref}

\begin{thebibliography}{10}
\providecommand{\url}[1]{#1}
\csname url@samestyle\endcsname
\providecommand{\newblock}{\relax}
\providecommand{\bibinfo}[2]{#2}
\providecommand{\BIBentrySTDinterwordspacing}{\spaceskip=0pt\relax}
\providecommand{\BIBentryALTinterwordstretchfactor}{4}
\providecommand{\BIBentryALTinterwordspacing}{\spaceskip=\fontdimen2\font plus
\BIBentryALTinterwordstretchfactor\fontdimen3\font minus
  \fontdimen4\font\relax}
\providecommand{\BIBforeignlanguage}[2]{{%
\expandafter\ifx\csname l@#1\endcsname\relax
\typeout{** WARNING: IEEEtran.bst: No hyphenation pattern has been}%
\typeout{** loaded for the language `#1'. Using the pattern for}%
\typeout{** the default language instead.}%
\else
\language=\csname l@#1\endcsname
\fi
#2}}
\providecommand{\BIBdecl}{\relax}
\BIBdecl

\bibitem{FCN}
E.~Shelhamer, J.~Long, and T.~Darrell, ``Fully convolutional networks for
  semantic segmentation,'' \emph{IEEE Trans. Pattern Anal. Mach. Intell.},
  vol.~39, no.~4, pp. 640--651, Apr. 2016.

\bibitem{Segnet}
V.~Badrinarayanan, A.~Kendall, and R.~Cipolla, ``Segnet: A deep convolutional
  encoder-decoder architecture for image segmentation,'' \emph{IEEE Trans.
  Pattern Anal. Mach. Intell.}, vol.~39, no.~12, pp. 2481--2495, Dec. 2017.

\bibitem{deeplab}
L.-C. Chen, G.~Papandreou, I.~Kokkinos, K.~Murphy, and A.~L. Yuille, ``Deeplab:
  Semantic image segmentation with deep convolutional nets, atrous convolution,
  and fully connected crfs,'' \emph{IEEE Trans. Pattern Anal. Mach. Intell.},
  vol.~40, no.~4, pp. 834--848, Apr. 2017.

\bibitem{stc}
Y.~Wei, X.~Liang, Y.~Chen, X.~Shen, M.-M. Cheng, J.~Feng, Y.~Zhao, and S.~Yan,
  ``Stc: A simple to complex framework for weakly-supervised semantic
  segmentation,'' \emph{IEEE Trans. Pattern Anal. Mach. Intell.}, vol.~39,
  no.~11, pp. 2314--2320, 2016.

\bibitem{A2GNN}
B.~Zhang, J.~Xiao, J.~Jiao, Y.~Wei, and Y.~Zhao, ``Affinity attention graph
  neural network for weakly supervised semantic segmentation,'' \emph{IEEE
  Trans. Pattern Anal. Mach. Intell.}, vol.~44, no.~11, pp. 8082--8096, 2021.

\bibitem{online}
P.-T. Jiang, L.-H. Han, Q.~Hou, M.-M. Cheng, and Y.~Wei, ``Online attention
  accumulation for weakly supervised semantic segmentation,'' \emph{IEEE Trans.
  Pattern Anal. Mach. Intell.}, vol.~44, no.~10, pp. 7062--7077, 2021.

\bibitem{LIID}
Y.~Liu, Y.-H. Wu, P.~Wen, Y.~Shi, Y.~Qiu, and M.-M. Cheng, ``Leveraging
  instance-, image-and dataset-level information for weakly supervised instance
  segmentation,'' \emph{IEEE Trans. Pattern Analy. Mach. Intell.}, vol.~44,
  no.~3, pp. 1415--1428, 2020.

\bibitem{salvage}
L.~Sui, C.-L. Zhang, and J.~Wu, ``Salvage of supervision in weakly supervised
  object detection and segmentation,'' \emph{IEEE Trans. Pattern Anal. Mach.
  Intell.}, 2023.

\bibitem{EPS++}
M.~Lee, S.~Lee, J.~Lee, and H.~Shim, ``Saliency as pseudo-pixel supervision for
  weakly and semi-supervised semantic segmentation,'' \emph{IEEE Trans. Pattern
  Anal. Mach. Intell.}, 2023.

\bibitem{MCIC}
J.~Fan and Z.~Zhang, ``Memory-based cross-image contexts for weakly supervised
  semantic segmentation,'' \emph{IEEE Trans. Pattern Anal. Mach. Intell.},
  2022.

\bibitem{MCIS}
W.~Wang, G.~Sun, and L.~Van~Gool, ``Looking beyond single images for weakly
  supervised semantic segmentation learning,'' \emph{IEEE Trans. Pattern Anal.
  Mach. Intell.}, 2022.

\bibitem{wei_revisiting}
Y.~Wei, H.~Xiao, H.~Shi, Z.~Jie, J.~Feng, and T.~S. Huang, ``Revisiting dilated
  convolution: A simple approach for weakly-and semi-supervised semantic
  segmentation,'' in \emph{Proc. IEEE Conf. Comput. Vis. Pattern Recognit.},
  2018, pp. 7268--7277.

\bibitem{whatpoint}
A.~Bearman, O.~Russakovsky, V.~Ferrari, and F.~F. Li, ``What's the {P}oint:
  Semantic segmentation with point supervision,'' in \emph{Eur. Conf. Comput.
  Vis.}, 2016.

\bibitem{kernel_cut}
M.~Tang, F.~Perazzi, A.~Djelouah, I.~Ben~Ayed, C.~Schroers, and Y.~Boykov, ``On
  regularized losses for weakly-supervised cnn segmentation,'' in \emph{Proc.
  Euro. Conf. Comput. Vis. (ECCV)}, 2018, pp. 507--522.

\bibitem{seminar}
H.~Chen, J.~Wang, H.~C. Chen, X.~Zhen, F.~Zheng, R.~Ji, and L.~Shao, ``Seminar
  learning for click-level weakly supervised semantic segmentation,'' in
  \emph{Proc. IEEE/CVF Int. Conf. Comput. Vis.}, 2021, pp. 6920--6929.

\bibitem{PSI}
J.~Xu, C.~Zhou, Z.~Cui, C.~Xu, Y.~Huang, P.~Shen, S.~Li, and J.~Yang,
  ``Scribble-supervised semantic segmentation inference,'' in \emph{Int. Conf.
  Comput. Vis.}, 2021, pp. 15\,354--15\,363.

\bibitem{Scribblesup}
D.~Lin, J.~Dai, J.~Jia, K.~He, and J.~Sun, ``Scribblesup: Scribble-supervised
  convolutional networks for semantic segmentation,'' in \emph{IEEE Conf.
  Comput. Vis. Pattern Recog.}, 2016, pp. 3159--3167.

\bibitem{URSS}
Z.~Pan, P.~Jiang, Y.~Wang, C.~Tu, and A.~G. Cohn, ``Scribble-supervised
  semantic segmentation by uncertainty reduction on neural representation and
  self-supervision on neural eigenspace,'' in \emph{Int. Conf. Comput. Vis.},
  2021, pp. 7416--7425.

\bibitem{boxsup}
J.~Dai, K.~He, and J.~Sun, ``Boxsup: Exploiting bounding boxes to supervise
  convolutional networks for semantic segmentation,'' in \emph{Int. Conf.
  Comput. Vis.}, 2015, pp. 1635--1643.

\bibitem{SDI}
A.~Khoreva, R.~Benenson, J.~Hosang, M.~Hein, and B.~Schiele, ``Simple does it:
  Weakly supervised instance and semantic segmentation,'' in \emph{IEEE Conf.
  Comput. Vis. Pattern Recog.}, 2017, pp. 876--885.

\bibitem{WSSL}
G.~Papandreou, L.-C. Chen, K.~P. Murphy, and A.~L. Yuille, ``Weakly-and
  semi-supervised learning of a deep convolutional network for semantic image
  segmentation,'' in \emph{Int. Conf. Comput. Vis.}, 2015, pp. 1742--1750.

\bibitem{cam}
B.~Zhou, A.~Khosla, A.~Lapedriza, A.~Oliva, and A.~Torralba, ``Learning deep
  features for discriminative localization,'' in \emph{IEEE Conf. Comput. Vis.
  Pattern Recog.}, 2016, pp. 2921--2929.

\bibitem{advcam}
J.~Lee, E.~Kim, and S.~Yoon, ``Anti-adversarially manipulated attributions for
  weakly and semi-supervised semantic segmentation,'' in \emph{Proc. IEEE Conf.
  Comput. Vis. Pattern Recognit.}, 2021, pp. 4071--4080.

\bibitem{t-sne}
L.~Van~der Maaten and G.~Hinton, ``Visualizing data using t-sne.'' \emph{Jour.
  Mach. Learn. Research}, vol.~9, no.~11, 2008.

\bibitem{singlestage}
N.~Araslanov and S.~Roth, ``Single-stage semantic segmentation from image
  labels,'' in \emph{IEEE Conf. Comput. Vis. Pattern Recog.}, 2020, pp.
  4253--4262.

\bibitem{random-walk}
P.~Vernaza and M.~Chandraker, ``Learning random-walk label propagation for
  weakly-supervised semantic segmentation,'' in \emph{IEEE Conf. Comput. Vis.
  Pattern Recog.}, 2017, pp. 7158--7166.

\bibitem{PSA}
J.~Ahn and S.~Kwak, ``Learning pixel-level semantic affinity with image-level
  supervision for weakly supervised semantic segmentation,'' in \emph{IEEE
  Conf. Comput. Vis. Pattern Recog.}, 2018, pp. 4981--4990.

\bibitem{crf}
P.~Kr{\"a}henb{\"u}hl and V.~Koltun, ``Parameter learning and convergent
  inference for dense random fields,'' in \emph{Inter. Conf. Mach.
  Learn.}\hskip 1em plus 0.5em minus 0.4em\relax PMLR, 2013, pp. 513--521.

\bibitem{survey}
W.~Shen, Z.~Peng, X.~Wang, H.~Wang, J.~Cen, D.~Jiang, L.~Xie, X.~Yang, and
  Q.~Tian, ``A survey on label-efficient deep image segmentation: Bridging the
  gap between weak supervision and dense prediction,'' \emph{EEE Trans. Pattern
  Anal. Mach. Intell.}, 2023.

\bibitem{AFA}
L.~Ru, Y.~Zhan, B.~Yu, and B.~Du, ``Learning affinity from attention:
  end-to-end weakly-supervised semantic segmentation with transformers,'' in
  \emph{IEEE Conf. Comput. Vis. Pattern Recog.}, 2022, pp. 16\,846--16\,855.

\bibitem{SPML}
T.-W. Ke, J.-J. Hwang, and S.~X. Yu, ``Universal weakly supervised segmentation
  by pixel-to-segment contrastive learning,'' in \emph{Inter. Conf. Learn.
  Repre.}, 2021.

\bibitem{IRN}
J.~Ahn, S.~Cho, and S.~Kwak, ``Weakly supervised learning of instance
  segmentation with inter-pixel relations,'' in \emph{IEEE Conf. Comput. Vis.
  Pattern Recog.}, 2019, pp. 2209--2218.

\bibitem{CISC_R}
L.~Wu, L.~Fang, X.~He, M.~He, J.~Ma, and Z.~Zhong, ``Querying labeled for
  unlabeled: Cross-image semantic consistency guided semi-supervised semantic
  segmentation,'' \emph{IEEE Trans. Pattern Anal. Mach. Intell.}, vol.~45,
  no.~7, pp. 8827--8844, Jul. 2023.

\bibitem{cian}
J.~Fan, Z.~Zhang, T.~Tan, C.~Song, and J.~Xiao, ``Cian: Cross-image affinity
  net for weakly supervised semantic segmentation,'' in \emph{Proc. AAAI Conf.
  Artif. Intell.}, vol.~34, no.~07, 2020, pp. 10\,762--10\,769.

\bibitem{l2g}
P.-T. Jiang, Y.~Yang, Q.~Hou, and Y.~Wei, ``L2g: A simple local-to-global
  knowledge transfer framework for weakly supervised semantic segmentation,''
  in \emph{IEEE Conf. Comput. Vis. Pattern Recog.}, 2022, pp. 16\,886--16\,896.

\bibitem{toco}
L.~Ru, H.~Zheng, Y.~Zhan, and B.~Du, ``Token contrast for weakly-supervised
  semantic segmentation,'' in \emph{IEEE Conf. Comput. Vis. Pattern Recog.},
  2023.

\bibitem{TEL}
Z.~Liang, T.~Wang, X.~Zhang, J.~Sun, and J.~Shen, ``Tree energy loss: Towards
  sparsely annotated semantic segmentation,'' in \emph{IEEE Conf. Comput. Vis.
  Pattern Recog.}, 2022, pp. 16\,907--16\,916.

\bibitem{normcut}
M.~Tang, A.~Djelouah, F.~Perazzi, Y.~Boykov, and C.~Schroers, ``Normalized cut
  loss for weakly-supervised cnn segmentation,'' in \emph{IEEE Conf. Comput.
  Vis. Pattern Recog.}, 2018, pp. 1818--1827.

\bibitem{GridCut}
D.~Marin, M.~Tang, I.~B. Ayed, and Y.~Boykov, ``Beyond gradient descent for
  regularized segmentation losses,'' in \emph{IEEE Conf. Comput. Vis. Pattern
  Recog.}, 2019, pp. 10\,187--10\,196.

\bibitem{AGMM}
L.~Wu, Z.~Zhong, L.~Fang, X.~He, Q.~Liu, J.~Ma, and H.~Chen, ``Sparsely
  annotated semantic segmentation with adaptive gaussian mixtures,'' in
  \emph{Proc. IEEE Conf. Comput. Vis. Pattern Recognit.}, 2023, pp.
  15\,454--15\,464.

\bibitem{CDL}
B.~Zhang, J.~Xiao, Y.~Wei, and Y.~Zhao, ``Credible dual-expert learning for
  weakly supervised semantic segmentation,'' \emph{Inter. Jour. Comput. Vis.},
  pp. 1--17, 2023.

\bibitem{resnet}
K.~He, X.~Zhang, S.~Ren, and J.~Sun, ``Deep residual learning for image
  recognition,'' in \emph{Proc. IEEE Conf. Comput Vis. Pattern Recognit.},
  2016, pp. 770--778.

\bibitem{vit}
A.~Dosovitskiy, L.~Beyer, A.~Kolesnikov, D.~Weissenborn, X.~Zhai,
  T.~Unterthiner, M.~Dehghani, M.~Minderer, G.~Heigold, S.~Gelly \emph{et~al.},
  ``An image is worth 16x16 words: Transformers for image recognition at
  scale,'' in \emph{Inter. Conf. Learn. Repres.}, 2020.

\bibitem{VOC}
M.~Everingham, L.~Van~Gool, C.~K. Williams, J.~Winn, and A.~Zisserman, ``The
  pascal visual object classes (voc) challenge,'' \emph{Int. Jour. Comput.
  Vision}, vol.~88, no.~2, pp. 303--338, Jun. 2010.

\bibitem{COCO}
T.-Y. Lin, M.~Maire, S.~Belongie, J.~Hays, P.~Perona, D.~Ramanan,
  P.~Doll{\'a}r, and C.~L. Zitnick, ``Microsoft coco: Common objects in
  context,'' in \emph{Euro Conf. Comput Vis.}\hskip 1em plus 0.5em minus
  0.4em\relax Springer, 2014, pp. 740--755.

\bibitem{city}
M.~Cordts, M.~Omran, S.~Ramos, T.~Rehfeld, M.~Enzweiler, R.~Benenson,
  U.~Franke, S.~Roth, and B.~Schiele, ``The cityscapes dataset for semantic
  urban scene understanding,'' in \emph{Proc. IEEE Conf. Comput. Vis. Pattern
  Recognit.}, 2016, pp. 3213--3223.

\bibitem{ade20k}
B.~Zhou, H.~Zhao, X.~Puig, S.~Fidler, A.~Barriuso, and A.~Torralba, ``Scene
  parsing through ade20k dataset,'' in \emph{IEEE Conf. Comput. Vis. Pattern
  Recognit.}, 2017, pp. 633--641.

\bibitem{seam}
Y.~Wang, J.~Zhang, M.~Kan, S.~Shan, and X.~Chen, ``Self-supervised equivariant
  attention mechanism for weakly supervised semantic segmentation,'' in
  \emph{IEEE Conf. Comput. Vis. Pattern Recog.}, 2020, pp. 12\,275--12\,284.

\bibitem{gradcam}
R.~R. Selvaraju, M.~Cogswell, A.~Das, R.~Vedantam, D.~Parikh, and D.~Batra,
  ``Grad-cam: Visual explanations from deep networks via gradient-based
  localization,'' in \emph{IEEE Int. Conf. Comput. Vis.}, 2017, pp. 618--626.

\bibitem{DSRG}
Z.~Huang, X.~Wang, J.~Wang, W.~Liu, and J.~Wang, ``Weakly-supervised semantic
  segmentation network with deep seeded region growing,'' in \emph{Proc. IEEE
  Conf. Comput. Vis. Pattern Recognit.}, 2018, pp. 7014--7023.

\bibitem{ficklenet}
J.~Lee, E.~Kim, S.~Lee, J.~Lee, and S.~Yoon, ``Ficklenet: Weakly and
  semi-supervised semantic image segmentation using stochastic inference,'' in
  \emph{Proc. IEEE Conf. Comput. Vis. Pattern Recognit.}, 2019, pp. 5267--5276.

\bibitem{DCA}
L.~Wu, M.~Lu, and L.~Fang, ``Deep covariance alignment for domain adaptive
  remote sensing image segmentation,'' \emph{IEEE Trans. Geosci. Remote Sens.},
  vol.~60, pp. 1--11, 2022.

\bibitem{qiang}
Q.~Liu, M.~He, Y.~Kuang, L.~Wu, J.~Yue, and L.~Fang, ``A multi-level
  label-aware semi-supervised framework for remote sensing scene
  classification,'' \emph{IEEE Trans. Geosci. Remote Sens.}, vol.~61, pp.
  1--12, 2023.

\bibitem{VoCo}
L.~Wu, J.~Zhuang, and H.~Chen, ``Voco: A simple-yet-effective volume
  contrastive learning framework for 3d medical image analysis,'' in \emph{IEEE
  Conf. Comput. Vis. Pattern Recog.}, 2024.

\bibitem{BPG}
B.~Wang, G.~Qi, S.~Tang, T.~Zhang, Y.~Wei, L.~Li, and Y.~Zhang, ``Boundary
  perception guidance: A scribble-supervised semantic segmentation approach,''
  in \emph{IJCAI Int. Joint Conf. Artifi. Intell.}, 2019.

\bibitem{gmm1}
M.~A. Ruzon and C.~Tomasi, ``Alpha estimation in natural images,'' in
  \emph{IEEE Conf. Comput. Vis. Pattern Recog. CVPR 2000}, vol.~1.\hskip 1em
  plus 0.5em minus 0.4em\relax IEEE, 2000, pp. 18--25.

\bibitem{gmm2}
Y.-Y. Chuang, B.~Curless, D.~H. Salesin, and R.~Szeliski, ``A bayesian approach
  to digital matting,'' in \emph{IEEE Conf. Comput. Vis. Pattern Recog. CVPR
  2001}, vol.~2.\hskip 1em plus 0.5em minus 0.4em\relax IEEE, 2001, pp. II--II.

\bibitem{gmm3}
Y.~Boykov and G.~Funka-Lea, ``Graph cuts and efficient nd image segmentation,''
  \emph{Int. J. Comput. Vis.}, vol.~70, no.~2, pp. 109--131, 2006.

\bibitem{gmm4}
C.~Rother, V.~Kolmogorov, and A.~Blake, ``Interactive foreground extraction
  using iterated graph cuts,'' \emph{ACM Trans. Graphics}, vol.~23, p.~3, 2012.

\bibitem{gmm5}
Y.~Boykov, O.~Veksler, and R.~Zabih, ``Fast approximate energy minimization via
  graph cuts,'' \emph{IEEE Trans. Pattern Anal. Mach. Intell.}, vol.~23,
  no.~11, pp. 1222--1239, 2001.

\bibitem{grabcut}
C.~Rother, V.~Kolmogorov, and A.~Blake, ``" grabcut" interactive foreground
  extraction using iterated graph cuts,'' \emph{ACM Trans. Graphics (TOG)},
  vol.~23, no.~3, pp. 309--314, 2004.

\bibitem{em1}
T.~K. Moon, ``The expectation-maximization algorithm,'' \emph{IEEE Signal
  Process. magazine}, vol.~13, no.~6, pp. 47--60, 1996.

\bibitem{em2}
C.~Liang, W.~Wang, J.~Miao, and Y.~Yang, ``Gmmseg: Gaussian mixture based
  generative semantic segmentation models,'' \emph{Adv. Neural Inform. Process.
  Syst.}, 2022.

\bibitem{em3}
Z.~Wang, S.~Wang, S.~Yang, H.~Li, J.~Li, and Z.~Li, ``Weakly supervised
  fine-grained image classification via gaussian mixture model oriented
  discriminative learning,'' in \emph{IEEE Conf. Comput. Vis. Pattern Recog.},
  2020, pp. 9749--9758.

\bibitem{TSCD}
R.~Xu, C.~Wang, J.~Sun, S.~Xu, W.~Meng, and X.~Zhang, ``Self correspondence
  distillation for end-to-end weakly-supervised semantic segmentation,'' in
  \emph{AAAI}, 2023.

\bibitem{RRM}
B.~Zhang, J.~Xiao, Y.~Wei, M.~Sun, and K.~Huang, ``Reliability does matter: An
  end-to-end weakly supervised semantic segmentation approach,'' in
  \emph{AAAI}, vol.~34, no.~07, 2020, pp. 12\,765--12\,772.

\bibitem{SLRNet}
J.~Pan, P.~Zhu, K.~Zhang, B.~Cao, Y.~Wang, D.~Zhang, J.~Han, and Q.~Hu,
  ``Learning self-supervised low-rank network for single-stage weakly and
  semi-supervised semantic segmentation,'' \emph{Inter. Jour. Comput. Vis.},
  vol. 130, no.~5, pp. 1181--1195, 2022.

\bibitem{OCCSE}
H.~Kweon, S.-H. Yoon, H.~Kim, D.~Park, and K.-J. Yoon, ``Unlocking the
  potential of ordinary classifier: Class-specific adversarial erasing
  framework for weakly supervised semantic segmentation,'' in \emph{Proc. IEEE
  Int. Conf. Comput. Vis.}, 2021, pp. 6994--7003.

\bibitem{CPN}
F.~Zhang, C.~Gu, C.~Zhang, and Y.~Dai, ``Complementary patch for weakly
  supervised semantic segmentation,'' in \emph{Proc. IEEE Int. Conf. Comput.
  Vis.}, 2021, pp. 7242--7251.

\bibitem{RIB}
J.~Lee, J.~Choi, J.~Mok, and S.~Yoon, ``Reducing information bottleneck for
  weakly supervised semantic segmentation,'' \emph{Adv. Neural Inform. Process.
  Sys.}, vol.~34, pp. 27\,408--27\,421, 2021.

\bibitem{VWL}
L.~Ru, B.~Du, Y.~Zhan, and C.~Wu, ``Weakly-supervised semantic segmentation
  with visual words learning and hybrid pooling,'' \emph{Inter. Jour. Comput.
  Vis.}, vol. 130, no.~4, pp. 1127--1144, 2022.

\bibitem{SIPE}
Q.~Chen, L.~Yang, J.-H. Lai, and X.~Xie, ``Self-supervised image-specific
  prototype exploration for weakly supervised semantic segmentation,'' in
  \emph{Proc. IEEE Conf. Comput. Vis. Pattern Recognit.}, 2022, pp. 4288--4298.

\bibitem{AMN}
J.~Lee, S.~J. Oh, S.~Yun, J.~Choe, E.~Kim, and S.~Yoon, ``Weakly supervised
  semantic segmentation using out-of-distribution data,'' in \emph{Proc. IEEE
  Conf. Comput. Vis. Pattern Recognit.}, 2022, pp. 16\,897--16\,906.

\bibitem{RECAM}
Z.~Chen, T.~Wang, X.~Wu, X.-S. Hua, H.~Zhang, and Q.~Sun, ``Class re-activation
  maps for weakly-supervised semantic segmentation,'' in \emph{Proc. IEEE Conf.
  Comput. Vis. Pattern Recognit.}, 2022, pp. 969--978.

\bibitem{ADELE}
S.~Liu, K.~Liu, W.~Zhu, Y.~Shen, and C.~Fernandez-Granda, ``Adaptive
  early-learning correction for segmentation from noisy annotations,'' in
  \emph{Proc. IEEE Conf. Comput. Vis. Pattern Recognit.}, 2022, pp. 2606--2616.

\bibitem{ESOL}
J.~LI, Z.~JIE, X.~Wang, L.~Ma \emph{et~al.}, ``Expansion and shrinkage of
  localization for weakly-supervised semantic segmentation,'' in \emph{Adv.
  Neural Inform. Process. Sys.}, 2022, pp. 1--12.

\bibitem{beco}
R.~Haifeng, B.~Tu, Z.~Wang, and J.~Li, ``Boundary-enhanced co-training for
  weakly supervised semantic segmentation,'' in \emph{Proc. IEEE Conf. Comput.
  Vis. Pattern Recognit.}, 2023.

\bibitem{conta}
D.~Zhang, H.~Zhang, J.~Tang, X.-S. Hua, and Q.~Sun, ``Causal intervention for
  weakly-supervised semantic segmentation,'' \emph{Adv. Neural Inform. Process.
  Sys.}, vol.~33, pp. 655--666, 2020.

\bibitem{EPS}
S.~Lee, M.~Lee, J.~Lee, and H.~Shim, ``Railroad is not a train: Saliency as
  pseudo-pixel supervision for weakly supervised semantic segmentation,'' in
  \emph{Proc. IEEE Conf. Comput. Vis. Pattern Recognit.}, 2021, pp. 5495--5505.

\bibitem{CDA}
Y.~Su, R.~Sun, G.~Lin, and Q.~Wu, ``Context decoupling augmentation for weakly
  supervised semantic segmentation,'' in \emph{Proc. IEEE Conf. Comput. Vis.
  Pattern Recognit.}, 2021, pp. 7004--7014.

\bibitem{WOOD}
J.~Lee, S.~J. Oh, S.~Yun, J.~Choe, E.~Kim, and S.~Yoon, ``Weakly supervised
  semantic segmentation using out-of-distribution data,'' in \emph{Proc. IEEE
  Conf. Comput. Vis. Pattern Recognit.}, 2022, pp. 16\,897--16\,906.

\bibitem{MCTFormer}
L.~Xu, W.~Ouyang, M.~Bennamoun, F.~Boussaid, and D.~Xu, ``Multi-class token
  transformer for weakly supervised semantic segmentation,'' in \emph{Proc.
  IEEE Conf. Comput. Vis. Pattern Recognit.}, 2022, pp. 4310--4319.

\bibitem{URN}
Y.~Li, Y.~Duan, Z.~Kuang, Y.~Chen, W.~Zhang, and X.~Li, ``Uncertainty
  estimation via response scaling for pseudo-mask noise mitigation in
  weakly-supervised semantic segmentation,'' in \emph{Proc. AAAI Conf. Artifi.
  Intell.}, vol.~36, no.~2, 2022, pp. 1447--1455.

\bibitem{DBFNet}
L.~Wu, L.~Fang, J.~Yue, B.~Zhang, P.~Ghamisi, and M.~He, ``Deep bilateral
  filtering network for point-supervised semantic segmentation in remote
  sensing images,'' \emph{IEEE Trans. Image Process.}, vol.~31, pp. 7419--7434,
  2022.

\bibitem{tree1}
L.~Song, Y.~Li, Z.~Li, G.~Yu, H.~Sun, J.~Sun, and N.~Zheng, ``Learnable tree
  filter for structure-preserving feature transform,'' \emph{Adv. Neural
  Inform. Process. Syst.}, vol.~32, 2019.

\bibitem{tree2}
L.~Song, Y.~Li, Z.~Jiang, Z.~Li, X.~Zhang, H.~Sun, J.~Sun, and N.~Zheng,
  ``Rethinking learnable tree filter for generic feature transform,''
  \emph{Adv. Neural Inform. Process. Syst.}, vol.~33, pp. 3991--4002, 2020.

\bibitem{tree3}
S.~Z. Li, ``Markov random field models in computer vision,'' in \emph{Eur.
  Conf. Comput. Vis.}\hskip 1em plus 0.5em minus 0.4em\relax Springer, 1994,
  pp. 361--370.

\bibitem{deeplabv3}
L.-C. Chen, Y.~Zhu, G.~Papandreou, F.~Schroff, and H.~Adam, ``Encoder-decoder
  with atrous separable convolution for semantic image segmentation,'' in
  \emph{Euro. Conf. Comput. Vis. (ECCV)}, 2018, pp. 801--818.

\bibitem{graphnet}
M.~Pu, Y.~Huang, Q.~Guan, and Q.~Zou, ``Graph{N}et: Learning image pseudo
  annotations for weakly-supervised semantic segmentation,'' in \emph{ACM Int.
  Conf. Multimedia}, 2018, pp. 483--491.

\bibitem{edge}
S.~Xie and Z.~Tu, ``Holistically-nested edge detection,'' in \emph{Int. Conf.
  Comput. Vis.}, 2015, pp. 1395--1403.

\bibitem{BCM}
C.~Song, Y.~Huang, W.~Ouyang, and L.~Wang, ``Box-driven class-wise region
  masking and filling rate guided loss for weakly supervised semantic
  segmentation,'' in \emph{Proc. IEEE Conf. Comput. Vis. Pattern Recognit.},
  2019, pp. 3136--3145.

\bibitem{bbam}
J.~Lee, J.~Yi, C.~Shin, and S.~Yoon, ``B{B}am: Bounding box attribution map for
  weakly supervised semantic and instance segmentation,'' in \emph{Proceedings
  of the IEEE/CVF conference on computer vision and pattern recognition}, 2021,
  pp. 2643--2652.

\bibitem{BAP}
Y.~Oh, B.~Kim, and B.~Ham, ``Background-aware pooling and noise-aware loss for
  weakly-supervised semantic segmentation,'' in \emph{Proc. IEEE Conf. Comput.
  Vis. Pattern Recognit.}, 2021, pp. 6913--6922.

\bibitem{Vit-pcm}
S.~Rossetti, D.~Zappia, M.~Sanzari, M.~Schaerf, and F.~Pirri, ``Max pooling
  with vision transformers reconciles class and shape in weakly supervised
  semantic segmentation,'' in \emph{ECCV}, 2022, pp. 446--463.

\bibitem{bayes}
X.~Wang, X.~Ma, and W.~E.~L. Grimson, ``Unsupervised activity perception in
  crowded and complicated scenes using hierarchical bayesian models,''
  \emph{IEEE Trans. Pattern Analy. Mach. Intell.}, vol.~31, no.~3, pp.
  539--555, 2008.

\end{thebibliography}
\bibliographystyle{IEEEtran}
\vspace{-.6in}

\end{document}